%% file: main.tex
\documentclass[pdftex,twocolumn,epjc3]{svjour3}          

\RequirePackage[T1]{fontenc}

\smartqed  

\RequirePackage{graphicx}
\usepackage{tabularx}
\usepackage{subcaption}
\RequirePackage{mathptmx}      
\RequirePackage{flushend}
\RequirePackage[colorlinks,citecolor=blue,urlcolor=blue,linkcolor=blue]{hyperref}
\usepackage{comment}
\usepackage{color,soul}
\RequirePackage{fix-cm}
\RequirePackage{amsmath}
\usepackage{mathtools}
\usepackage{isotope}
\RequirePackage{amssymb}
\RequirePackage{multirow}
\RequirePackage{hyperref}
\RequirePackage{xspace}
\RequirePackage{adjustbox}
\RequirePackage{rotating}
\RequirePackage{microtype}
\RequirePackage[compat=1.1.0]{tikz-feynman}
\tikzfeynmanset{warn luatex=false}
\RequirePackage{tikz}
\usetikzlibrary{decorations.pathreplacing,calligraphy}
\usetikzlibrary{snakes,shapes.geometric}
\tikzstyle{terminator} = [rectangle, draw, text centered, rounded corners, minimum height=2em]
\tikzstyle{process} = [rectangle, draw, text centered, minimum height=2em]
\input{newcommands.tex}

\usepackage[
    bibstyle=ieee,
    citestyle=numeric-comp,
    sorting=none,
    date=year,
    doi=true,
    isbn=false,
    eprint=true,
    url=true,
    maxnames=3, minnames=1,
]{biblatex}
\usepackage{lineno}

\journalname{Eur. Phys. J. C}

\begin{document}
\title{The M\textsc{onument} Experiment: Ordinary Muon Capture studies for \onbb decay}


\author{G.~R.~Araujo\thanksref{addr1}\and
        D.~Bajpai \thanksref{addr2} \and
        L.~Baudis \thanksref{addr1} \and
        V.~Belov \thanksref{addr4, addr5} \and
        E.~Bossio \thanksref{addr5, addr6, e2} \and
        T.E.~Cocolios \thanksref{addr7} \and
        H.~Ejiri \thanksref{addr8} \and
        M.~Fomina \thanksref{addr4} \and
        K.~Gusev \thanksref{addr4, addr5} \and
        I.H.~Hashim \thanksref{addr9} \and
        M.~Heines \thanksref{addr7} \and
        S.~Kazartsev \thanksref{addr4} \and
        A.~Knecht \thanksref{addr10} \and
        E.~Mondrag\'on \thanksref{addr5, e1} \and
        Z.W.~Ng \thanksref{addr9} \and
        I.~Ostrovskiy \thanksref{addr2, addr3} \and
        F.~Othman \thanksref{addr9} \and
        N.~Rumyantseva \thanksref{addr4, addr5} \and
        S.~Sch\"{o}nert \thanksref{addr5} \and
        M.~Schwarz \thanksref{addr5} \and
        E.~Shevchik \thanksref{addr4} \and
        M.~Shirchenko \thanksref{addr4} \and
        Yu.~Shitov \thanksref{addr11} \and
        E. O. Sushenok \thanksref{addr4} \and
        J.~Suhonen \thanksref{addr12, addr13} \and
        S.M.~Vogiatzi \thanksref{addr10} \and
        C.~Wiesinger \thanksref{addr5} \and
        I.~Zhitnikov \thanksref{addr4} \and
        D.~Zinatulina \thanksref{addr4} \and
}

\thankstext{e1}{e-mail: elizabeth.mondragon@tum.de}
\thankstext{e2}{e-mail: elisabetta.bossio@cea.fr}

\institute{Physik-Institut, University of Zurich, Zurich, Switzerland\label{addr1}
          \and
          Department of Physics and Astronomy, University of Alabama, Tuscaloosa, AL, USA\label{addr2}
          \and
          \emph{Present Address:}Institute of High Energy Physics, Beijing, China\label{addr3}
          \and
          Joint Institute for Nuclear Research, Dubna, Russia\label{addr4}
          \and
          Technische Universit\"{a}t M\"{u}nchen, Garching, Germany\label{addr5}
          \and
          \emph{Present Address:} IRFU, CEA, Université Paris-Saclay, Gif-sur-Yvette, France\label{addr6}
          \and
          KU Leuven, Institute for Nuclear and Radiation Physics, Leuven,~Belgium\label{addr7}
          \and
          Research Center on Nuclear Physics, Osaka University, Ibaraki, Osaka,~Japan\label{addr8}
          \and
          Department of Physics, Faculty of Science, Universiti Teknologi Malaysia, Johor Bahru, Malaysia\label{addr9}
          \and
          Paul Scherrer Institut, Villigen, Switzerland\label{addr10}
          \and
          Institute of Experimental and Applied Physics, Czech Technical University in Prague, Prague, Czech Republic\label{addr11}
          \and
          Department of Physics, University of  Jyv\"{a}skyl\"{a}, Jyv\"{a}skyl\"{a}, Finland\label{addr12}
          \and
          International Centre for Advanced Training and Research in Physics (CIFRA), Bucharest-Magurele, Romania\label{addr13}
}

\date{Received: date / Accepted: date}

\maketitle
\begin{abstract}\label{sec:abstract}
    The M\textsc{onument} experiment measures ordinary muon capture (OMC) on isotopes relevant for neutrinoless double-beta (\onbb) decay and nuclear astrophysics. 
    OMC is a particularly attractive tool for improving the theoretical description of \onbb decay. 
    It involves similar momentum transfers and allows testing the virtual transitions involved in \onbb decay against experimental data. 
    During the 2021 campaign, M\textsc{onument} measured OMC on $^{76}$Se and $^{136}$Ba, the isotopes relevant for next-generation \onbb decay searches, like L\textsc{egend} and n\textsc{EXO}.
    The experimental setup has been designed to accurately extract the total and partial muon capture rates, which requires precise reconstruction of energies and time-dependent intensities of the OMC-related $\gamma$ rays. 
    The setup also includes a veto counter system to allow selecting a clean sample of OMC events.
    This work provides a detailed description of the M\textsc{onument} setup operated during the 2021 campaign, its two DAQ systems, calibration and analysis approaches, and summarises the achieved detector performance.
    Future improvements are also discussed.
\end{abstract}

\input{1-motivation}
\input{2-approach}
\input{3-experiment}
\input{4-procedures}
\input{5-performances}
\input{6-conclusions}
\input{7-Acknowledgements}

\begingroup
\setlength{\emergencystretch}{2em}
\printbibliography

\end{document}

%% file: newcommands.tex
\newcommand{\Beta}    {\ensuremath{\beta}\xspace}

\newcommand{\onbb}    {\ensuremath{0\nu\beta\beta}\xspace}

\newcommand{\onHL}    {\ensuremath{T_{1/2}^{0\nu}}\xspace}

\newcommand{\qbb}     {\ensuremath{Q_{\beta\beta}}\xspace}

\newcommand{\mbb}     {m\ensuremath{_{\beta\beta}}\xspace}

%% file: 1-motivation.tex
\section{Motivation}\label{sec:motivation}
One of the most sensitive ways to determine whether neutrinos are Majorana fermions is to search for the neutrinoless double-beta (\onbb) decay. 
This process violates lepton number conservation, which is forbidden in the standard model of particle physics.
If it exists, it would prove that the neutrino Majorana mass contribution is non-vanishing regardless of the decay mechanism~\cite{Schechter:1982}. 

If the dominant mechanism is the exchange of a light Majorana neutrino, then the half-life T$_{1/2}$ depends on the effective Majorana mass $\mbb$ as follows: 
\begin{equation*}\label{eq:onbb_decay_rate}
    \frac{1}{\onHL} = \mathcal{G}^{0\nu}(\qbb,Z) \, g_A^4\, |\mathcal{M}^{0\nu}|^2 \, \frac{\mbb^2}{m_e^2} \;,
\end{equation*}
where $\mathcal{G}^{0\nu}(\qbb,Z)$ is the phase-space factor, $g_A$ is the axial-vector coupling constant, $\mathcal{M}^{0\nu}$ is the nuclear matrix element (NME), and $m_e$ is the electron mass.
Next-generation experiments~\cite{nexo,legend,cupid} will aim to probe effective Majorana masses as low as $\sim$10-20 meV.
However, the final sensitivity of next-generation experiments to $\mbb$ depends on the values of the NMEs and, for phenomenological nuclear models, on how the effective value of g$_A$ is renormalised to account for an incomplete description of the underlying nuclear structure~\cite{suhonen_2017,Engel_2017}.
A better understanding of these issues is one of the priorities of nuclear theory. 
It would reduce the ambiguity in the reach of the currently planned experiments and help guide the design of future efforts.

\begin{figure}[t!]
\begin{centering} 
\includegraphics[width=0.93\columnwidth]{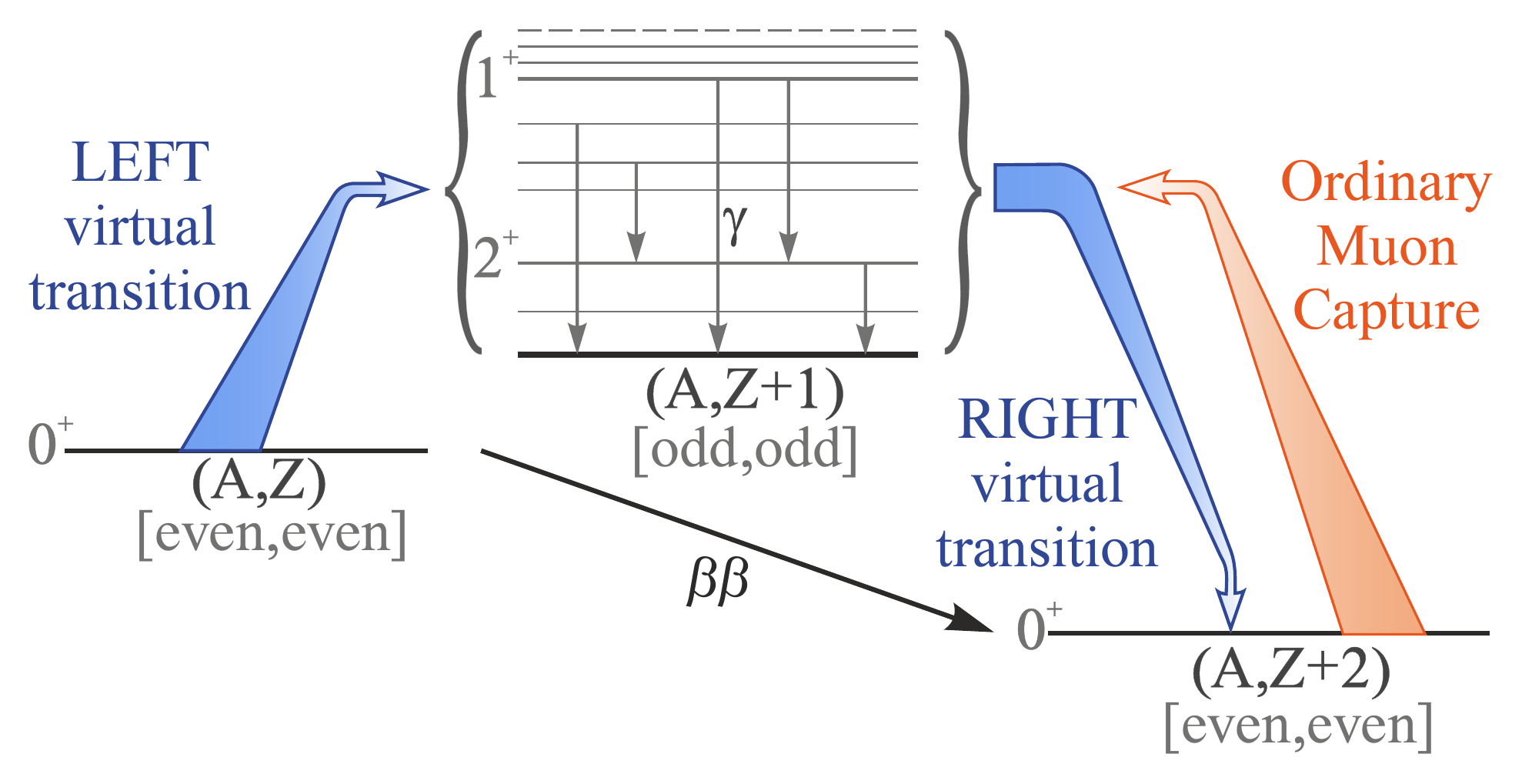}
\caption{Schematic representation of double-\Beta decay as a virtual transition through the ex states of the intermediate nucleus. 
In the \onbb decay, high-multipole states J$^\pi$ of the intermediate nucleus are involved. 
OMC on a target consisting of the (A, Z+2) daughter nucleus can provide information about the right virtual transition, accessing the same excited states of the intermediate nucleus.}
\label{fig:level-shceme}   
\end{centering}
\end{figure}

Benchmarking the nuclear structure calculations of the \onbb decay NMEs is possible, for example, with the measured half-lives of single-$\beta$ and two-neutrino double-$\beta$ decays~\cite{Suhonen:1998ck, Engel_2017, Suhonen:2004pw} or with single-particle transfer reactions~\cite{Freeman:2012hr, Kay:2008dg}. 
However, these processes involve momentum exchanges of several MeV, which is more than an order of magnitude lower than in \onbb decay. 
As was suggested in Ref.~\cite{Kortelainen:2002bz,Kortelainen2003}, ordinary muon capture (OMC) could be more advantageous for benchmarking the \onbb decay NMEs. 
OMC refers to the non-radiative capture of a negative muon from the atomic orbital. 
This process results in momentum exchanges of 50-100 MeV, leading to highly excited final states with high multipolarity, similar to the case of \onbb decay. 
Fig.~\ref{fig:level-shceme} shows a schematic diagram of the \onbb decay, with the OMC effectively representing the right ($\beta^{+}$) branch of the \onbb decay virtual transitions. 
An additional motivation for studying the OMC is provided by its ability to probe nuclear structure calculations relevant for solar and supernova neutrinos~\cite{Pinedo2014, Balasi2015}. 
M\textsc{onument} is an experiment dedicated to measuring OMC on isotopes relevant for \onbb decay in addition to some OMC astrophysical motivated measurements. 
It builds upon the approach demonstrated by earlier work~\cite{dubna_omc_2019}. 
One of its primary aims is providing the input necessary for a systematic study of the sensitivity of the OMC strength function to the effective values of the weak axial couplings, which should help improve the accuracy of the \onbb decay NME calculations~\cite{suhonen_omc_2020}. 
M\textsc{onument}'s first measurement campaign was performed at the Paul Scherrer Institute (PSI), Switzerland, in 2021. 
The OMC was measured on $^{76}$Se and $^{136}$Ba isotopically enriched targets, which are particularly relevant for the next-generation \onbb decay searches with L\textsc{egend}~\cite{legend} and n\textsc{EXO}~\cite{nexo}. 
This work presents a detailed description of the experimental setup used during this campaign, its two data acquisition (DAQ) systems, as well as the calibration and analysis strategies. 
It summarises the achieved performance and discusses improvements anticipated for the upcoming measurement campaigns. 
 The results of the total and partial OMC capture rates will be reported elsewhere.

%% file: 2-approach.tex
\section{Measurement Approach}\label{sec:approach}
The OMC on a $\isotope[A][Z+2]{Z}$ nucleus populates several excited states of a daughter ($\isotope[A][Z+1]{Y^{*}}$) nucleus. 
The main goal of the experiment is to measure the partial capture rates, or relative yields, to different excited states of the daughter nucleus. 
These rates provide the information on the nuclear structure needed to advance the nuclear physics models' calculations. 

Extracting the partial rates can be done by measuring and comparing the intensities of the $\gamma$ rays de-exciting the relevant levels. 
Measuring the time evolution of the de-excitation $\gamma$ rays also allows one to extract the total capture rate, which corresponds to the lifetime of the muonic atom.

The capture of a muon is always preceded by a cascade of muonic-X rays ($\mu$X rays) corresponding to the de-excitation of the formed muonic atom. 
The intensity of these $\mu$X-ray lines can therefore give information on the number of muons stopped in the target.

\begin{figure}[t!]
    \begin{centering} 
    \includegraphics[width=0.85\columnwidth]{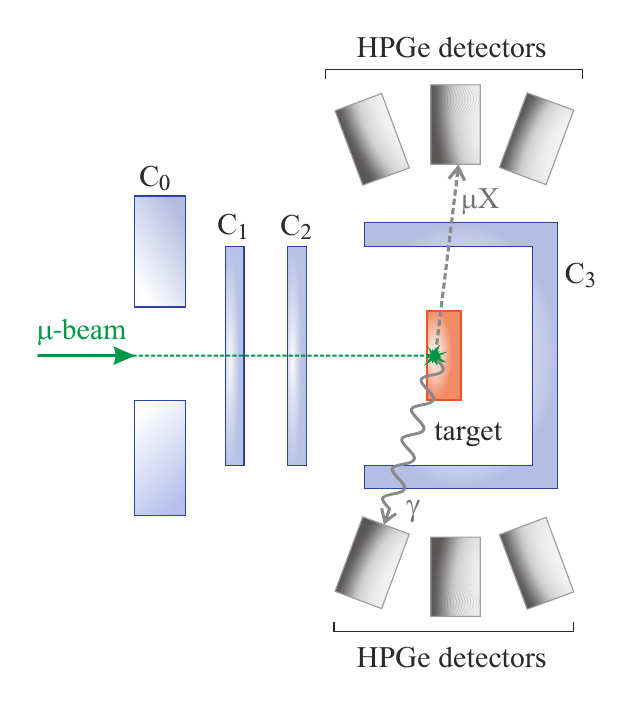}
    \caption{Schematic view of the measurement. 
        It consists of four scintillator counters used to tag muons from the beam that are stopped in the target: a ring-shaped veto counter C$_0$, two thin pass-through counters C$_1$ and C$_2$ that are placed before the target, and a cup-like veto counter C$_3$ surrounding the target. 
        HPGe detectors placed around the target measure the \textit{prompt} $\mu$X rays and the \textit{delayed} $\gamma$ rays.}
    \label{fig:setup_sketch}   
    \end{centering}
    \end{figure}

The measurement principle is sketched in Fig.~\ref{fig:setup_sketch}.
A designated target is exposed to a muon beam, with adjustments made to its momentum and position to optimise the likelihood of observing OMC.
Four scintillator counters surround the target: a ring-shaped veto counter C$_0$, two thin pass-through counters C$_1$ and C$_2$, and a cup-like veto counter C$_3$. 
They allow defining the trigger for OMC:

\begin{equation}\label{eq:trigger_omc}
    \mu_{\mathrm{stop}} = \overline{\textrm C}_0 \wedge {\textrm C}_1 \wedge {\textrm C}_2 \wedge \overline{\textrm C}_3 \;.
\end{equation}

The anti-coincidence of C$_0$ and coincidence of C$_1$ and C$_2$ allow selecting only muons coming from the beam and hitting the target. 
The additional anti-coincidence of C$_3$ ensures the muon was stopped in the target. 
Further elaboration on the precise application of this trigger condition is provided in Sec.~\ref{sec:trigger-rates}.

The setup is completed with an array of High-Purity Germanium (HPGe) detectors placed around the target.
The HPGe detectors are used to measure the time and energy distribution of the $\gamma$ and $\mu$X rays.

Events that satisfy the trigger condition, i.e., the signals in the HPGe detectors that occur within a defined time window after a muon stops in the target, are classified as correlated events.
Details on the time window to build correlated events are given later in Sec.~\ref{sec:trigger-rates}.
These events are mostly attributed to the $\mu$X and $\gamma$ rays emitted after OMC and can be identified in the energy spectrum by their characteristic energy. 
The time information of the event allows to further separate $\mu$X from $\gamma$ rays. 
While the first are expected within picoseconds from the stopped muon, the emission of $\gamma$ rays extends for hundreds of nanoseconds after the muon capture, depending on the characteristic lifetime of the muonic atom. 
Among correlated events, the firsts are classified as \textit{prompt} events and the seconds as \textit{delayed} events.

Events that do not pass the trigger condition are classified as \textit{uncorrelated} events. 
These events are attributed to different processes that constitute the background of the experiment.

Lastly, events recorded during the short periods of time when the beam is not operational are classified as \textit{beam-off} events. 
The energy spectrum of these events contains decays with intermediate lifetimes, too long to be seen in the energy spectrum of correlated events but too short to be seen hours after the irradiation.
Examples of the energy spectra for the different classes of events will be given in Sec.~\ref{sec:trigger-rates} and Sec.~\ref{sec:conclusions}.

After the measurements with the muon beam, the targets are placed in a separate screening station to measure the long-lived activity following OMC. 
The measurements performed outside the beam hall are referred to as the ``offline measurements'' and are only briefly discussed in Sec.~\ref{sec:offline-mes}.

%% file: 3-experiment.tex
\section{Experiment}\label{sec:experiment}
The experiment was conducted at the $\pi$E1 beam line of the PSI, which meets the experimental prerequisites for muon energy and beam intensity. 
The beam is composed of negatively charged muons~\cite{psi}. 
The muons' momentum was set to approximately 40~MeV/c, with a beam dispersion of roughly 2\%. 
The beam's intensity was estimated to be around 10$^4$ muons per second, with a corresponding rate of stopped muons in the target of a similar magnitude (see Sec. \ref{sec:trigger-rates}).

The 2021 campaign consisted of the measurement of $^{136}$Ba, $^{\textrm{nat}}$Ba, $^{76}$Se and $^{\textrm{nat}}$Se over three weeks between October and November.
These are referred to as \textit{physics} runs.
Two measurements with $^{136}$Ba, chronologically named $^{136}$Ba-I and $^{136}$Ba-II, were performed, one at the beginning, after the $^{\textrm{nat}}$Ba measurement, and one at the end of the campaign, after the completion of the $^{\textrm{nat}}$Se and $^{76}$Se measurements.
Before and after each physics run, calibration measurements were performed using $\gamma$-ray sources $^{152}$Eu, $^{60}$Co, $^{88}$Y, $^{133}$Ba and a $^{\textrm{nat}}$Pb OMC target.
These runs are referred to as \textit{calibration} runs and are named chronologically from I to IV.
Unlike all the other calibration sources, the $^{\textrm{nat}}$Pb source was exposed to the muon beam to produce high-energy $\mu$X rays with energies up to about 6 MeV.

The HPGe detector array (Sec. \ref{sec:HPGe-detectors}) in this campaign included eight HPGe detectors mounted on an aluminium frame and surrounding the target chamber (Sec. \ref{sec:target-chamber}) at a distance of $\sim$15\,cm.
Another HPGe detector was positioned beneath the detector array to monitor the background.

The target chamber is composed of a target holder, surrounded by the four muon counters, placed to trace the path of incoming muons towards the target, as described in Sec. \ref{sec:approach}.

Two parallel DAQ systems -- MIDAS and ALPACA -- were employed and are described in Sec. \ref{sec:daq}. 
\subsection{Target Chamber}\label{sec:target-chamber}
\begin{figure}[t!]
\begin{centering} 
\includegraphics[width=1.0\columnwidth]{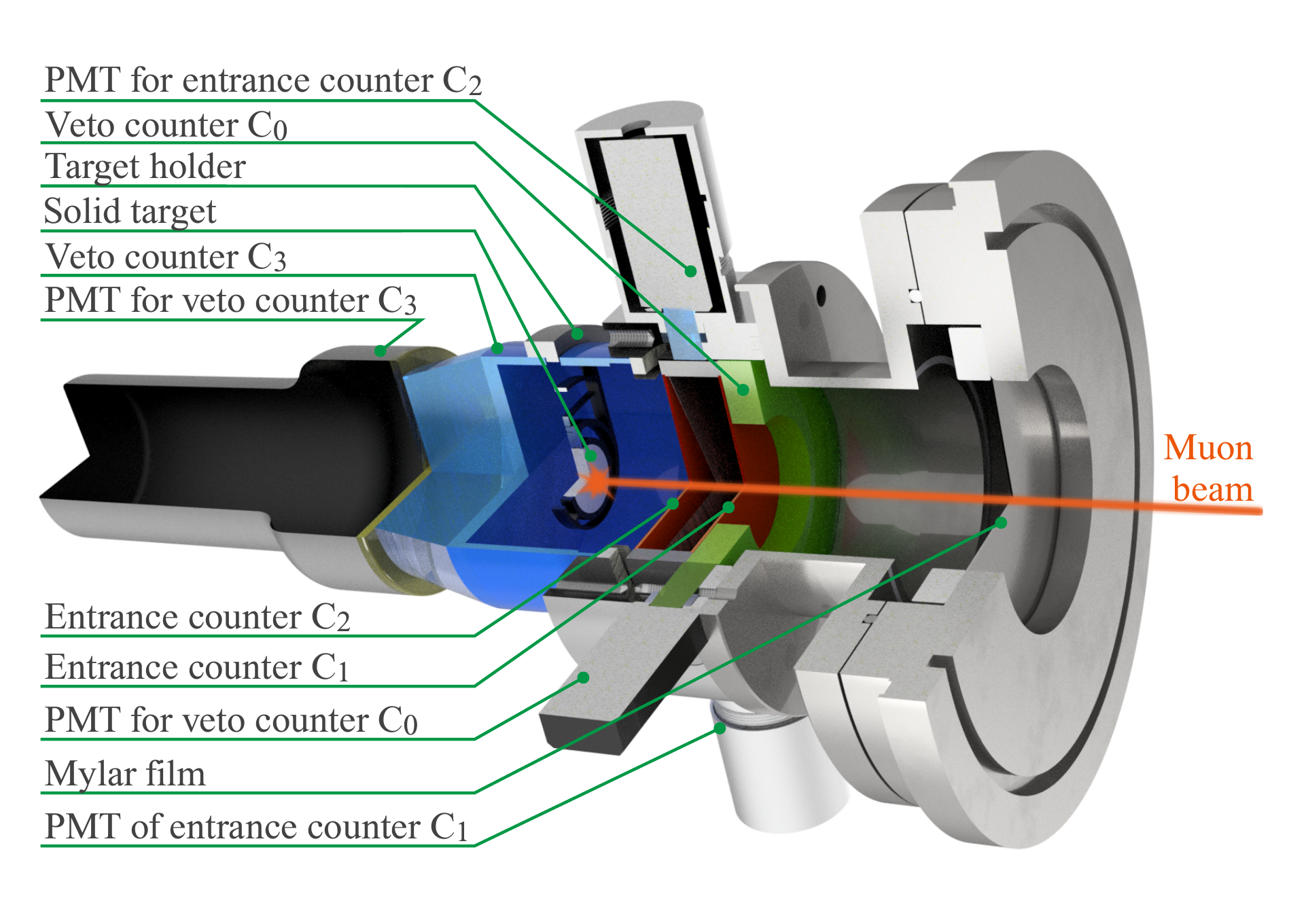}
\caption{Schematic view of the target chamber.}
\label{fig:target-chamber}   
\end{centering}
\end{figure}

The four muon counters comprising the target chamber are made of scintillating material and connected to photomultiplier tubes (PMTs), as shown in Fig.~\ref{fig:target-chamber}. 
C$_0$, located at the entrance of the target enclosure, is a 1-centimeter thick ring-shaped scintillator.
The C$_1$ and C$_2$ pass-through counters have a thickness of 0.5\,mm each.
Finally, C$_3$ is the cup-shaped counter that surrounds the target. 
All counters are made of low-Z polystyrene scintillating material attached to a light guide.
The targets were processed and sized in-house to fit in the designated holder.
Elemental selenium powder was pressed to form a 2 g, 20 mm in diameter and 1.8 mm in thickness tablet.
This target mass was chosen to provide sufficient rates of muon capture (see Sec. \ref{sec:trigger-rates}).

For the barium targets, powders of $^{{\textrm{nat}}}$BaCO$_3$ and $^{136}$BaCO$_3$ salts were acquired.
Barium carbonate solutions in isopropanol were deposited onto plastic holders and let to dry.
Once dried, the holders were sealed with lids and fixed with Kapton-tape.
Table \ref{tab:target-abundance} lists all the targets used in the present study.
\begin{table}[b]
\caption{Targets used in the 2021 campaign. This table includes their abundance, composition, mass and thickness.}
\begin{tabular}[h]{c c c c c}
\hline
Target & Enrichment & Composition & Mass & Thickness\\
& ($\%$) & & (g) & (mm) \\
\hline
$^{{\textrm{nat}}}$Se &  -- &  metal powder & 2.0 & 1.8 \\
$^{{76}}$Se  &  99.97 & metal powder &  2.0 &  1.8 \\
$^{{\textrm{nat}}}$Ba &  -- &  BaCO$_3$ powder &  2.0 &  2.4 \\
$^{136}$Ba   &  95.27 &  BaCO$_3$ powder&  2.0 & 2.4 \\
\hline
\end{tabular}
\label{tab:target-abundance}
\end{table}
\subsection{Germanium Detector Array}\label{sec:HPGe-detectors}
\begin{figure}[t!]
  \begin{centering} 
  \includegraphics[width=0.9\columnwidth]{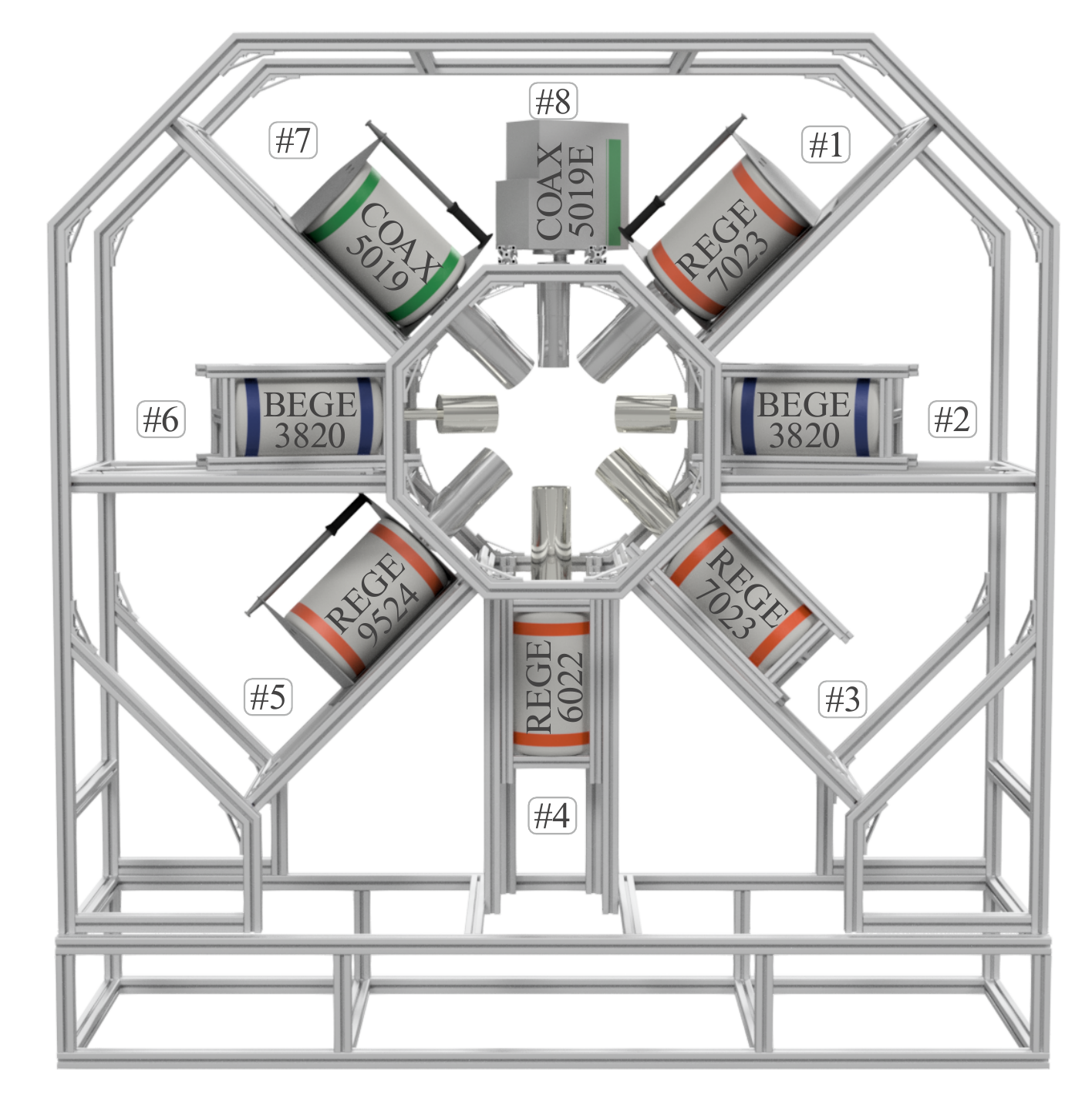}
  \caption{Schematic view of the aluminium frame with the eight HPGe detectors used in the 2021 measurement campaign.
  It consisted of a pair of BEGe detectors in positions $\#$2 and $\#$6 with the same efficiency and six coaxial detectors. 
  The p-type COAX detectors with the same efficiency were in positions $\#$7 and $\#$8; all other coaxial detectors were n-type REGe and had different detection efficiencies.
  }
  \label{fig:GeDet}   
  \end{centering}
  \end{figure}

The HPGe detector array's holder frame was designed and manufactured in-house using standard industrial 40x40~mm$^{2}$ aluminium-alloy profiles, as seen in Fig.~\ref{fig:GeDet}.

\begin{figure*}[t!]
  \centering
  \begin{subfigure}[t]{0.32\textwidth}
    \includegraphics[height=8cm]{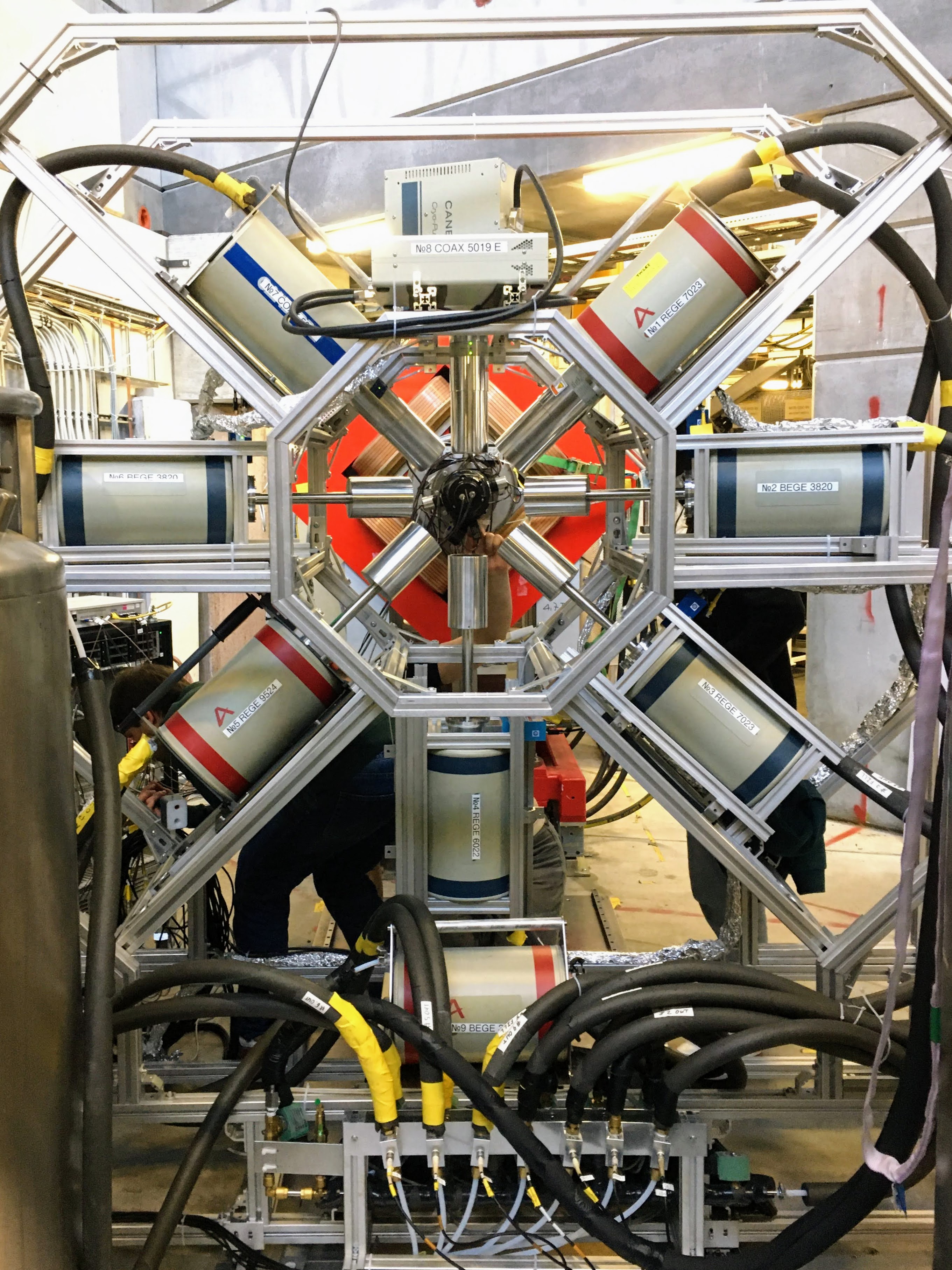}
    \label{fig:exp-setup1}
  \end{subfigure}
  \hfill
  \begin{subfigure}[t]{0.62\textwidth}
    \includegraphics[height=8cm]{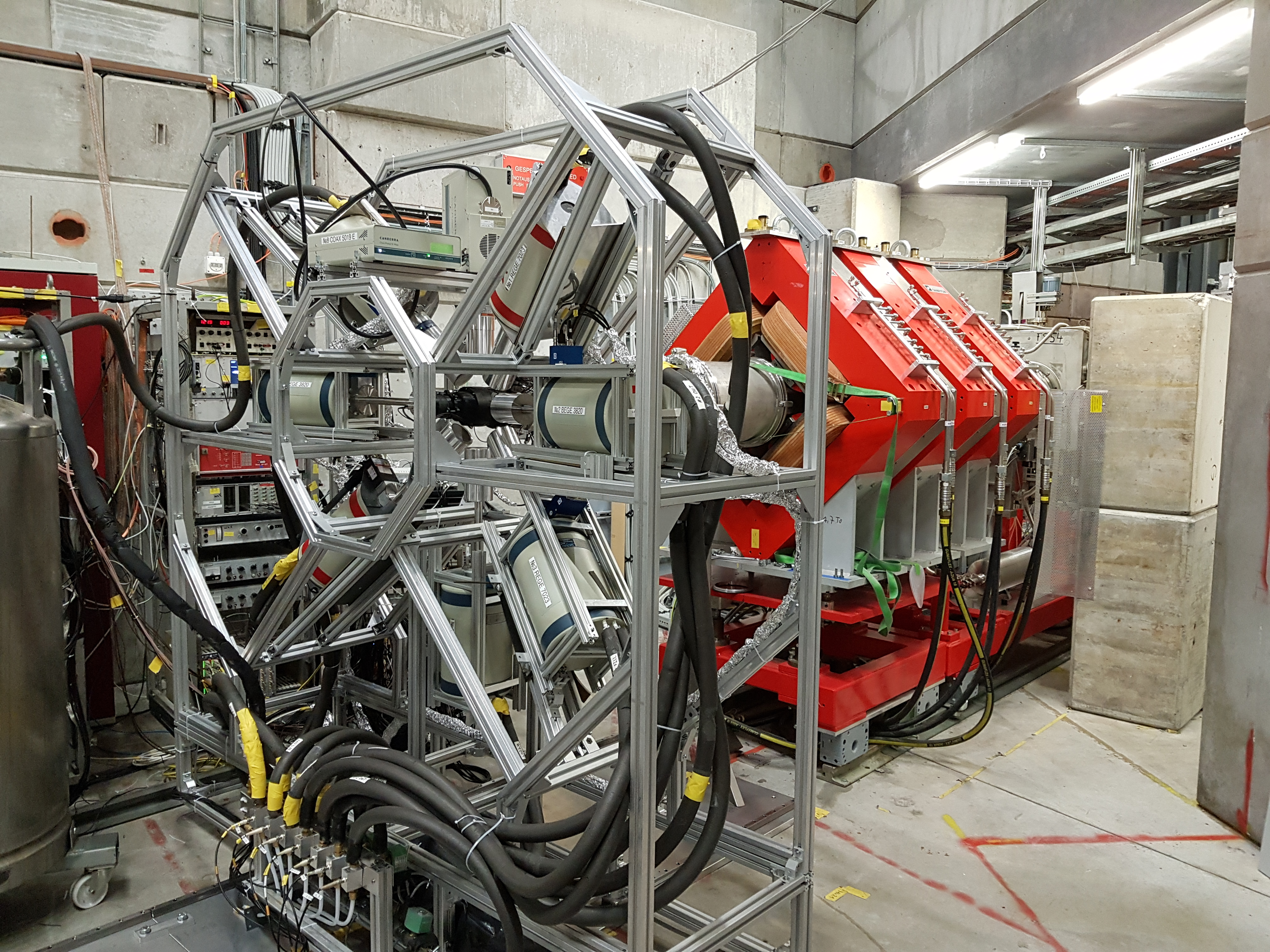}
    \label{fig:exp-setup2}
  \end{subfigure}
  \caption{
    Photos of the experimental setup. 
    They show the HPGe detector array surrounding the target chamber.
  }
  \label{fig:array}
\end{figure*}

The eight HPGe detectors mounted on the frame consisted of two large-volume p-type coaxial (\textbf{COAX}) detectors, four large-volume n-type coaxial reverse-electrode (\textbf{REGe}) detectors, and two p-type broad energy (\textbf{BEGe}) detectors. The REGe and BEGe detectors had thin beryllium windows. 
Another BEGe detector, dedicated to monitoring the background, was placed at the bottom of the frame.
All detectors were mounted in Big-MAC cryostats except for one of the COAX detectors, which was mounted in a Cryo-Pulse electrically refrigerated cryostat. 
A laser alignment tool was employed to set their precise locations around the target chamber at the beamline's end.
The auto-filling system at PSI (similar to the one described in~\cite{refilling}) facilitated the automatic filling of liquid nitrogen in most of the detectors every three days.
Fig.~\ref{fig:array} shows the array.
The refilling system installed at the bottom of the frame is also visible.

\subsection{Electronics}\label{sec:electronics}
\begin{figure*}[t!]
  \begin{centering} 
  \includegraphics[width=0.8\textwidth]{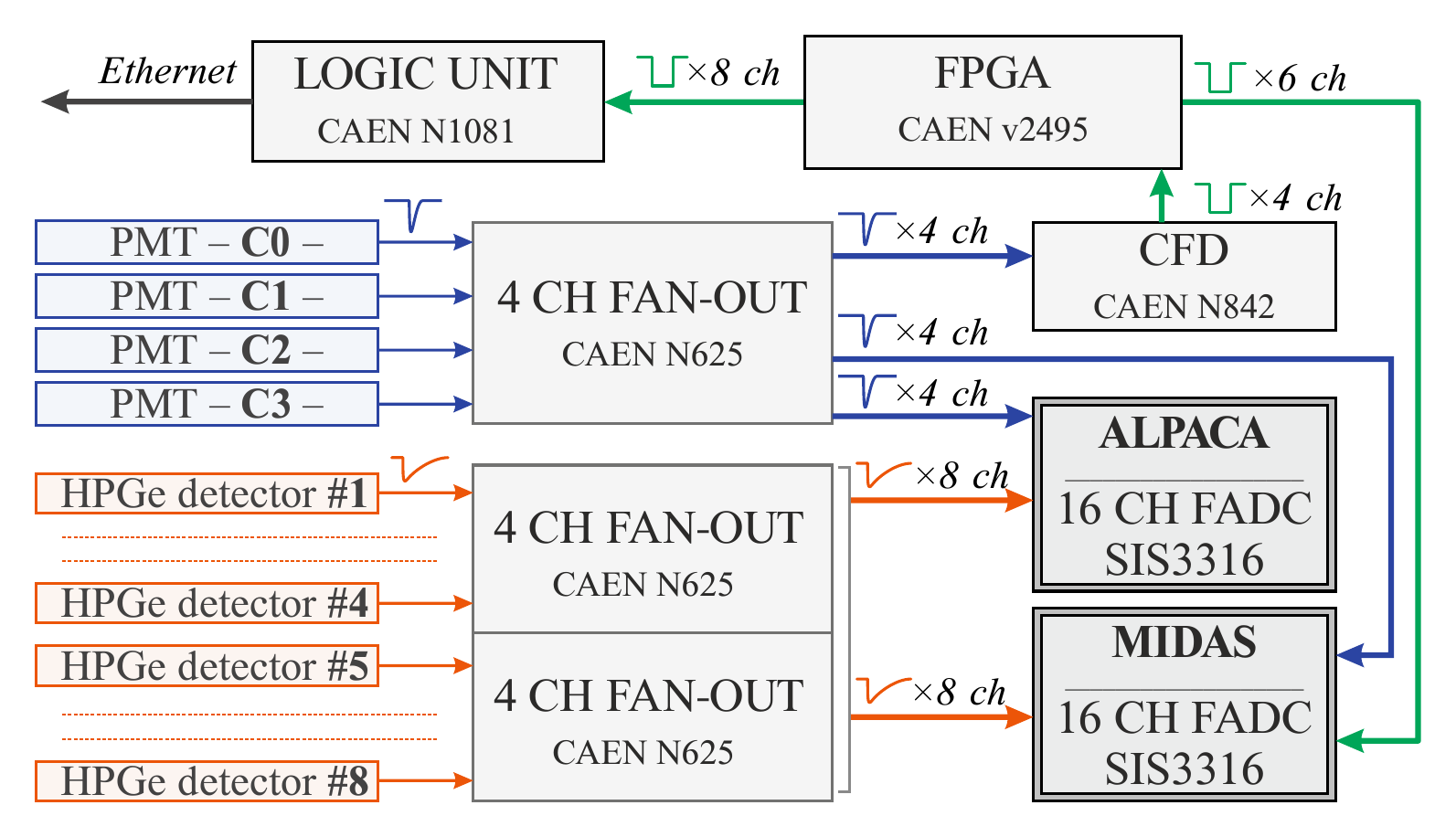}
  \caption{
  Electronic scheme from the 2021 measurement campaign.
  The output from all detectors (PMTs and HPGe) is fed to a fan-in fan-out module, which then sends a parallel signal to ALPACA and MIDAS.
  Additionally, PMT signals are also sent to a logic unit to monitor the counters' rates during the measurements. 
  }
  \label{fig:electronics}   
  \end{centering}
  \end{figure*}
Fig.~\ref{fig:electronics} shows the overview of the electronic configuration used during the 2021 measurement campaign.
The PMTs' analogue outputs were connected to the front-end unit, which consisted of a shaping amplifier and a discriminator. 
The amplifier outputs were subsequently directed to fan-in fan-out units (CAEN N625 NIM) to feed the two used DAQ systems, as described in the next section. 
The logic signals from the discriminator were employed to fine-tune the beam and ensure data-quality monitoring. 
To accomplish this, a CAEN V2495 programmable logic unit was utilised. 
This unit can configure any hardware trigger scheme through onboard FPGA programming.
A second CAEN N1081A programmable logic unit was employed to monitor the count rate in each counter. 
The high voltage of the PMTs was monitored and adjusted by industrial Digital Analogue Convertors (DACs) and specialized control software. 
These devices facilitate the remote setting of the PMTs' high voltage, hardware thresholds, and logic scheme, which is critical to avoid disrupting the beam or halting the data acquisition run.
The firmware was optimised for regulating pileups. 
\subsection{Data acquisition systems} \label{sec:daq}
The two DAQ systems employed in 2021 -- ALPACA and MIDAS -- were independent and operated in parallel. 
ALPACA~\cite{mario} is a homemade system originally designed for a rare-event experiment and adapted for M\textsc{onument}.
MIDAS~\cite{midas:daq} is a PSI and TRIUMF's development and is especially suited for handling the large amounts of data typical from accelerator experiments without requiring a fast storage server. 

These two systems complement each other by gathering information via distinct methods yet ensuring comprehensive data for a complete analysis.
The two software programs ran on two Struck SIS-3316 digitiser modules (FADCs), each with 16 channels. 
MIDAS used a 14-bit resolution model, which provides a sampling frequency of 250~MHz, while ALPACA used a 16-bit resolution model, with a sampling frequency of 125\,MHz, which offers waveform-compression features.
These compression features were exploited and needed in ALPACA. 
\subsubsection{MIDAS}\label{sec:MIDAS}
The two systems differ primarily in their trigger scheme and data recording methods.
MIDAS relies on online digital signal processing (DSP) performed by the FADC. 

In MIDAS, each detector -- HPGe detectors and PMTs -- is triggered independently using individual trigger thresholds.
This scheme provides flexibility when looking for delayed coincidences. 
A dead time of 1.4 $\mu$s follows each trigger. 
The online DSP is performed during this time, and the relevant information is collected. 
MIDAS does not record events during this time window but sets a pileup protection flag in case a trigger is issued during the period.

The waveforms from the HPGe detectors undergo initial processing through a trapezoidal filter, followed by the extraction and subsequent storage of energy and trigger time information onto a disk.
The FADC also extracts the trigger time and signal amplitude for the PMT waveforms online, stored for the following analysis. 
Additionally, a 1.4 $\mu$s-long waveform centred at the rising edge is saved for HPGe detectors exclusively, which can help improve the time resolution in offline analysis.
\subsubsection{ALPACA}\label{sec:ALPACA}
ALPACA's implementation is built on source code provided by Struck Innovative Systeme, which includes the Ethernet interface class and basic code specific to the SIS-3316 device.

Due to the original framework for which the ALPACA software was developed, ALPACA does not use online processing on the FADC.
Instead, signal waveforms from the HPGe detectors and PMTs are stored, enabling offline time and energy reconstruction.

The FADC is connected via a CAT-6 Ethernet cable to a dedicated readout server hosting hard drives in RAID-6 configuration.
The digitised data -- waveforms -- are first sampled from the analogue signals on the FADC and stored in its memory.
Simultaneously, the ALPACA program running on the server requests the transfer of previously sampled data from the FADC's memory to a RAM buffer on the server.
The program periodically writes the contents of the buffer to consecutive files.

In detail, the memory of the FADC is organised in two banks. 
One ``active'' bank receives sampled data while the content of the other ``passive'' bank is transferred to the server at the same time.
Each bank is subdivided into 16 buffers, each sized 64 MB and dedicated to a single input channel.
Every individual channel buffer is first filled with sampled waveforms.
Once at least a small part of the buffers is filled and the readout of the ``passive'' bank is finished, the software issues a ``bank swap'' command, swapping the role of both banks and emptying the new active bank.
A bank swap always affects all channel buffers simultaneously.
The contents of the new passive bank are subsequently transferred to the server.
These technical details on the storage will become paramount when discussing the dead time of the system in Sec.~\ref{sec:livetime}.

The triggering scheme in ALPACA operates as follows: when an energy deposition occurs in one of the HPGe detectors, that detector and all the PMTs are read out.
Other HPGe detectors are not read out during the event recording, greatly reducing data throughput.
Avoiding individual PMT pulse triggering effectively mitigates systematic dead times.

\begin{figure*}[t!]
    \centering
    \includegraphics[width=0.9\textwidth]{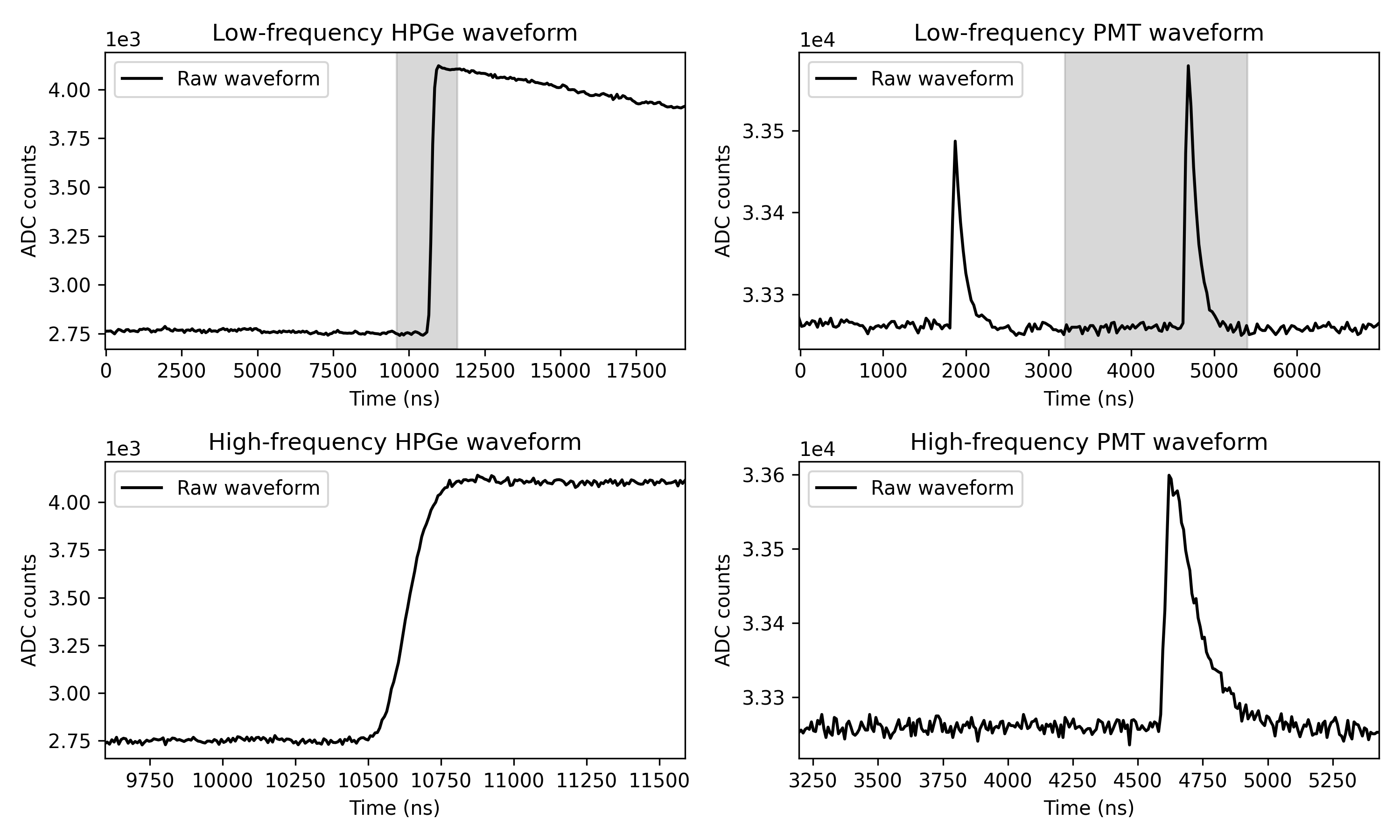}
    \caption{
    Acquired waveforms using ALPACA, with the HPGe detectors' traces on the left and the PMTs' on the right. 
    The low-frequency waveforms are shown at the top, with the grey-shaded region indicating the time window during which the waveform is also recorded in high-frequency mode (bottom).
    }
    \label{fig:alpaca_traces}
\end{figure*}

\begin{figure*}[t!]
  \centering
  \begin{tikzpicture}
      
      \draw(2.5,6) node [left, black, text width=1cm, font=\large] {ALPACA};
      \draw(9.5,6) node [left, black, text width=1cm, font=\large] {MIDAS};
  
      \draw[thick] (0,0) node [left] {Ge 1} -- (5,0);
      \draw[thick] (0,1.5) node [left] {Ge 2} -- (5,1.5);
      \draw[thick] (0,3) node [left] {PMT 1} -- (5,3);
      \draw[thick] (0,4.5) node [left] {PMT 2} -- (5,4.5);
      
      \draw[thick] (1,1.5) -- (1,1.9) to [out=90, in=180] (2,1.5);
      \draw[fill=red,red] (1,1.4) -- (1.1,1.2) node [right, black, text width=1cm, font=\tiny] {Trigger 1 (primary)} -- (0.9,1.2) -- cycle;
      \draw[fill=red,red] (1,2.9) -- (1.1,2.7) node [right, black, text width=1cm, font=\tiny] {Trigger 1 (induced)} -- (0.9,2.7) -- cycle;
      \draw[fill=red,red] (1,4.4) -- (1.1,4.2) node [right, black, text width=1cm, font=\tiny] {Trigger 1 (induced)} -- (0.9,4.2) -- cycle;
      \node[rectangle,fill=red!80!yellow!70, anchor=south west, minimum height=0.9cm, minimum width=1.3cm, opacity=0.3] at (0.7,1.4) {}; 
      
      \draw[thick] (3.5,0) -- (3.5,0.4) to [out=90, in=180] (4.5,0);
      \draw[fill=red,red] (3.5,-0.1) -- (3.6,-0.3) node [right, black, text width=1cm, font=\tiny] {Trigger 2 (primary)} -- (3.4,-0.3) -- cycle;
      \draw[fill=red,red] (3.5,2.9) -- (3.6,2.7) node [right, black, text width=1cm, font=\tiny] {Trigger 2 (induced)} -- (3.4,2.7) -- cycle;
      \draw[fill=red,red] (3.5,4.4) -- (3.6,4.2) node [right, black, text width=1cm, font=\tiny] {Trigger 2 (induced)} -- (3.4,4.2) -- cycle;
      \node[rectangle,fill=red!80!yellow!70, anchor=south west, minimum height=0.9cm, minimum width=1.3cm, opacity=0.3] at (3.2,-0.1) {};
      
      \draw[thick] (0.8,3) -- (0.8,3.4);
      \node[rectangle,fill=red!80!yellow!70, anchor=south west, minimum height=0.9cm, minimum width=1.cm, opacity=0.3] at (0.3,2.9) {};
      \draw[thick] (1.8,3) -- (1.8,3.6);
      \draw[thick] (3,3) -- (3,3.5);
      \node[rectangle,fill=red!80!yellow!70, anchor=south west, minimum height=0.9cm, minimum width=1.cm, opacity=0.3] at (2.8,2.9) {};
      \draw[thick] (3.3,3) -- (3.3,3.3);
      \draw[thick] (4.7,3) -- (4.7,3.3);
      \draw[thick] (0.5,4.5) -- (0.5,5.2);
      \node[rectangle,fill=red!80!yellow!70, anchor=south west, minimum height=0.9cm, minimum width=1.cm, opacity=0.3] at (0.3,4.4) {};
      \draw[thick] (1.7,4.5) -- (1.7,5.);
      \draw[thick] (2.9,4.5) -- (2.9,4.9);
      \node[rectangle,fill=red!80!yellow!70, anchor=south west, minimum height=0.9cm, minimum width=1.cm, opacity=0.3] at (2.8,4.4) {};
      \draw[thick] (4.8,4.5) -- (4.8,5.);
      
      \draw[thick] (7,0) node [left] {Ge 1} -- (12,0); 
      \draw[thick] (7,1.5) node [left] {Ge 2} -- (12,1.5); 
      \draw[thick] (7,3) node [left] {PMT 1} -- (12,3); 
      \draw[thick] (7,4.5) node [left] {PMT 2} -- (12,4.5); 
      
      \draw[thick] (8,1.5) -- (8,1.9) to [out=90, in=180] (9,1.5);
      \node[rectangle,fill=gray!50, anchor=south west, minimum height=0.9cm, minimum width=0.55cm, opacity=0.3] at (7.8,1.4) {};
      \draw[fill=red,red] (8,1.4) -- (8.1,1.2) node [right, black, text width=1.2cm, font=\tiny] {Trigger 1 (Ge)} -- (7.9,1.2) -- cycle;
      \node[rectangle,fill=red!80!yellow!70, anchor=south west, minimum height=0.9cm, minimum width=0.25cm, opacity=0.3] at (7.9,1.4) {};
      
      \draw[thick] (10.5,0) -- (10.5,0.4) to [out=90, in=180] (11.5,0);
      \node[rectangle,fill=gray!50, anchor=south west, minimum height=0.9cm, minimum width=0.55cm, opacity=0.3] at (10.3,-0.1) {};
      \draw[fill=red,red] (10.5, -0.10) -- (10.6,-0.3) node [right, black, text width=1.2cm, font=\tiny] {Trigger 2 (Ge)} -- (10.4,-0.3) -- cycle;
      \node[rectangle,fill=red!80!yellow!70, anchor=south west, minimum height=0.9cm, minimum width=0.25cm, opacity=0.3] at (10.4,-0.1) {};
      
      \draw[thick] (7.8,3) -- (7.8,3.4); 
      \draw[fill=red,red] (7.8,2.9) -- (7.9,2.7) node [right, black, text width=0.8cm, font=\tiny] {Trigger 2 (PMT)} -- (7.7,2.7) -- cycle;
      \node[rectangle,fill=gray!50, anchor=south west, minimum height=0.9cm, minimum width=0.55cm, opacity=0.3] at (7.6,2.9) {};
      \draw[thick] (8.8,3) -- (8.8,3.6);
      \draw[fill=red,red] (8.8,2.9) -- (8.9,2.7) node [right, black, text width=0.8cm, font=\tiny] {Trigger 4 (PMT)} -- (8.7,2.7) -- cycle;
      \node[rectangle,fill=gray!50, anchor=south west, minimum height=0.9cm, minimum width=0.55cm, opacity=0.3] at (8.6,2.9) {};
      \draw[thick] (10,3) -- (10,3.5);
      \draw[fill=red,red] (10,2.9) -- (9.9,2.7) -- (10.1,2.7) node [right, black, text width=0.8cm, font=\tiny] {Trigger 6 (PMT)} -- cycle;
      \node[rectangle,fill=gray!50, anchor=south west, minimum height=0.9cm, minimum width=0.55cm, opacity=0.3] at (9.8,2.9) {};
      \draw[thick] (10.3,3) -- (10.3,3.3);
      \draw[thick] (11.7,3) -- (11.7,3.3);
      \draw[fill=red,red] (11.7,2.9) -- (11.6,2.7) -- (11.8,2.7) node [right, black, text width=0.8cm, font=\tiny] {Trigger 7 (PMT)} -- cycle;
      \node[rectangle,fill=gray!50, anchor=south west, minimum height=0.9cm, minimum width=0.55cm, opacity=0.3] at (11.5,2.9) {};
      
      \draw[thick] (7.5,4.5) -- (7.5,5.2);
      \draw[fill=red,red] (7.5,4.4) -- (7.6,4.2) node [right, black, text width=0.8cm, font=\tiny] {Trigger 1 (PMT)} -- (7.4,4.2)-- cycle;
      \node[rectangle,fill=gray!50, anchor=south west, minimum height=0.9cm, minimum width=0.55cm, opacity=0.3] at (7.3,4.4) {};
      \draw[thick] (8.7,4.5) -- (8.7,5.);
      \draw[fill=red,red] (8.7,4.4) -- (8.8,4.2) node [right, black, text width=0.8cm, font=\tiny] {Trigger 3 (PMT)} -- (8.6,4.2) -- cycle;
      \node[rectangle,fill=gray!50, anchor=south west, minimum height=0.9cm, minimum width=0.55cm, opacity=0.3] at (8.5,4.4) {};      
      \draw[thick] (9.9,4.5) -- (9.9,4.9);
      \draw[fill=red,red] (9.9,4.4) -- (10.,4.2) node [right, black, text width=0.8cm, font=\tiny] {Trigger 5 (PMT)} -- (9.8,4.2) -- cycle;
      \node[rectangle,fill=gray!50, anchor=south west, minimum height=0.9cm, minimum width=0.55cm, opacity=0.3] at (9.7,4.4) {};
      \draw[thick] (11.8,4.5) -- (11.8,5.);
      \draw[fill=red,red] (11.8,4.4) -- (11.9,4.2) node [right, black, text width=0.8cm, font=\tiny] {Trigger 8 (PMT)} -- (11.7,4.2) -- cycle;
      \node[rectangle,fill=gray!50, anchor=south west, minimum height=0.9cm, minimum width=0.55cm, opacity=0.3] at (11.6,4.4) {};
  \end{tikzpicture}
  \caption{
    Comparison of the trigger schemes employed by the ALPACA (left) and MIDAS (right) DAQ systems, using only two HPGe detectors and two PMTs, to simplify the concept. 
    In ALPACA, when a HPGe detector triggers (red triangles), its trace and those from all PMT detectors are recorded.
    The complete traces are recorded for offline analysis (red area).
    On the other hand, MIDAS triggers each HPGe and PMT detectors separately. 
    The FADC reconstructs each signal's trigger time and energy online. 
    Some important information may be lost during the online reconstruction window (grey areas), as illustrated in the PMT 1 trace after Trigger 6, where the DAQ cannot register the following signal.
    This is a deadtime window that lasts 1.4 $\mu$s.
    Only a short sample from the HPGe detectors' trace is recorded for offline analysis (red area). 
    }
  \label{fig:daq_llama_midas}
\end{figure*}
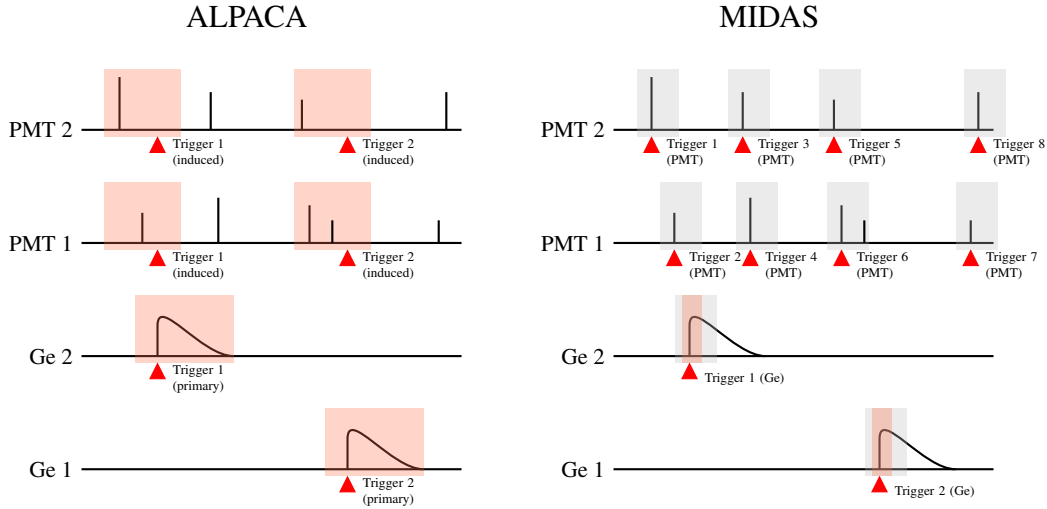

The following explanation delves into the benefits of utilizing the 16-bit FADC model's compression features.
Given the expected high event rate and data throughput, a long trace with the highest sampling frequency would quickly saturate ALPACA.
A compromise was met by recording a low-frequency long trace (20 $\mu$s at 15\,MHz) for the HPGe detectors' events.
This trace suffices to do a good offline energy reconstruction and optimise the energy resolution of these events (see Sec.\,\ref{sec:dsp}).
For optimal timing resolution, a shorter trace centred around the leading edge of the waveform is recorded at the maximum sampling frequency (2 $\mu$s at 125\,MHz).
In the case of the PMTs, a 2 $\mu$s high-frequency trace is recorded about 2 $\mu$s prior to the trigger. 
This maximises the resolution of the trigger time of a muon registered in a PMT in relation to the trigger time of $\mu$X and $\gamma$ rays registered by the HPGe detector following OMC in the target.
Additionally, a longer 30 MHz low-frequency PMT trace of about 7 $\mu$s (5.4 $\mu$s before the trigger and 1.6 $\mu$s after) is recorded to detect delayed coincidences. 
An example of the recorded HPGe detector and PMT waveforms is shown in Fig.~\ref{fig:alpaca_traces}.
The two different trigger schemes from MIDAS and ALPACA are exemplified in Fig. \ref{fig:daq_llama_midas}.

\subsection{Offline Measurements}\label{sec:offline-mes}
The delayed $\gamma$ rays following OMC in $^{76}$Se and $^{136}$Ba have half-lives of a few milliseconds to years.
To detect the activity of the delayed $\gamma$ rays associated with the neutron, proton, and heavy charged particle unbound processes after OMC reactions, the targets were moved to an offline measurement setup after about one week of exposure to the muon beam and measured for 21 days. 
The setup consisted of a HPGe n-type detector mounted in a vertical dipstick cryostat inside a lead shield provided by the PSI radiochemistry department. 
The signals from the detector were digitised using a \textsc{Canberra} ADC Genie-2000 basic spectroscopy software.
The summed spectrum was saved every hour to track the time evolution of $\gamma$ rays in the energy spectrum. 

%% file: 4-procedures.tex
\section{Analysis Procedures}\label{sec:procedures}
\subsection{MIDAS}
A MIDAS DAQ event contains the uncalibrated energy after applying a trapezoidal filter, the trigger time and DSP-related information (baseline, pileup, overflow, underflow, etc.) for each of the detectors. 
The HPGe events additionally contain the 1.4 $\mu$s waveforms.
The analysis chain comprises several steps:
\begin{itemize}
\item[1.-] Data format conversion from the MIDAS storage structure to ROOT format without any cuts or selections.

\item[2.-] Data quality cuts based on the beam stability, target type, etc.

\item[3.-] Energy calibration for the HPGe-detectors.

\item[4.-] Production of the correlated and uncorrelated energy spectra.
\end{itemize}

A more detailed description of the MIDAS DSP, similar to the one used for this work, is provided in~\cite{Skawran}.
\subsection{ALPACA}
The analysis chain for ALPACA involves data selection and also production steps. 
Data selection involves removing data taken under unstable experimental conditions, such as detector refilling, beam tuning, and changes in setup. 
The list of accepted files is then stored and used for further steps of the analysis.
As shown on the left panel of Fig.~\ref{fig:daq_llama_midas}, an event in ALPACA typically comprises one germanium waveform and four PMT waveforms. 
The DSP of these waveforms is conducted offline using tools from the \textsc{GERDA} experiment~\cite{Agostini:2011xe}.
ALPACA data is organised in a tier-like structure, where relevant information is extracted and condensed through successive analysis steps:
\begin{itemize}
    \item[1.-] Tier~1 involves transforming binary data into ROOT format, including all waveforms and auxiliary FADC parameters required for subsequent analysis, like the events' timestamps.
    
    \item[2.-] Tier~2 contains the outcome of DSP performed on Tier~1 data providing specific event information. 
    This includes detector ID, event's trigger position and amplitude.
    The energy deposited in a HPGe detector is reconstructed via an optimised trapezoidal filter, while PMTs' DSP is optimised for timing information (Sec.~\ref{sec:dsp}). 
    
    \item[3.-]Tier~3 contains calibrated event information, including parameters derived from applying calibration and quality cuts criteria (Sec.~\ref{sec:qcs}). 
    It also includes parameters combining information from HPGe detectors and PMTs, such as multiplicity (the number of coincident HPGe detector signals) and the time difference between PMT and HPGe detector signals. 
    
    \item[4.-]Tier~4 is the final stage of ALPACA's data structure, where events are classified as correlated, prompt, delayed, or uncorrelated. 
    This last classification is necessary for the high-level analysis.
\end{itemize}
ALPACA is discussed in further extent due to the novelty of the system and the fact that it is the in-house DAQ.
\subsubsection{Digital Signal Processing }\label{sec:dsp}
The DSP of the HPGe detectors’ waveforms starts with the evaluation of the low-frequency trace’s (Fig.~\ref{fig:alpaca_traces}, top-left) first 8 $\mu$s. 
For this, the baseline's mean value, RMS, and exponential behaviour are considered.

The energy of the event was reconstructed with an optimized trapezoidal filter.
The trapezoidal filter consists of a combination of a Moving Window Deconvolution (MWD) and a Moving Window Average (MWA), whose sizes are related to the rise time (RT) and the flat top (FT) of the resulting trapezoidal pulse, with MWA $=$ RT and MWD $=$ RT+FT~\cite{STEIN1996141}. 
The amplitude of the trapezoidal filter at a fixed position in the FT delivers the energy value. 
Several combinations of values for the MWD and MWA were tested to deduce which provides the best energy resolution. 
A general preference for long shaping parameters was observed for all the detectors. 
As a result, a MWD size of 8 $\mu$s and a MWA size of 5 $\mu$s were chosen for all the HPGe detectors. 
The optimized trapezoidal filter allows for obtaining a better energy resolution than the MIDAS DAQ, as will be shown in Table~\ref{tab:resolution}.

The trigger position was determined using a leading-edge trigger algorithm after applying a fast-trapezoidal filter (MWD $=$ MWA $=$ 384 ns) on the waveforms. 
This filter allows identifying multiple triggers in the same traces that can be discarded as pileup events. 
To identify non-physical events, parameters such as the minimum and maximum position in the trace and pulse rise time were also evaluated.

While most DSP algorithms were applied to the low-frequency waveform, the trigger position and rise time were evaluated on the high-frequency waveform (Fig.~\ref{fig:alpaca_traces}, bottom-left) to obtain better time resolution.
The DSP of the PMT traces consists of a pulse finding algorithm. 
First, the baseline and RMS are calculated recursively, excluding every time frame of the trace that exceeds 4$\times$RMS of the previously determined mean value. 
The different pulses are then identified by a simple leading-edge trigger algorithm and their trigger positions and pulse amplitudes evaluated. 
Because of a long shaping time of the PMT pulses, a fast MWD filter (MWD $=$ 8 ns) was applied to the PMT traces before the above pulse finding algorithm.
This allowed for an enhanced resolution of multiple pulses in the same trace in close proximity.

\subsubsection{Quality cuts}\label{sec:qcs}
For the ALPACA analysis, a set of data quality cuts was developed to select only the events where the waveforms resemble a single HPGe detector pulse.
Quality cuts typically reject non-physical signals, pileup events, and signals exceeding the FADC dynamic range.
For this study, exclusively the HPGe detector low-frequency waveforms were used, given they are significantly longer than the high-frequency ones but preserve sufficiently precise trigger information.

\begin{figure}[t!]
\begin{centering} 
\includegraphics[width=1\columnwidth]{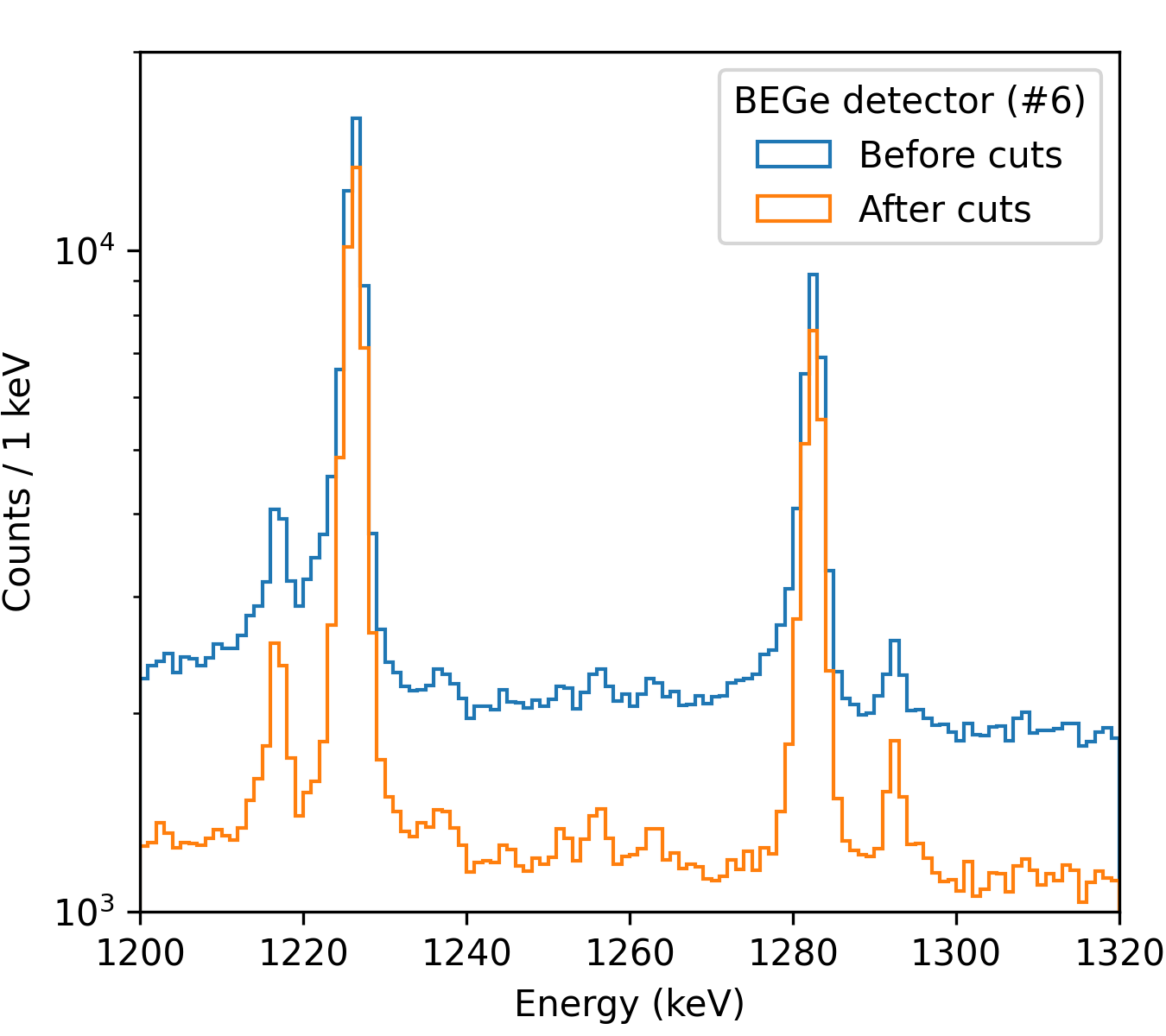}
\caption{
$^{136}$Ba L-series $\mu$X rays before (blue) and after (orange) applying quality cuts to the ALPACA data for the BEGe detector $\#$6.
One can see how the $\mu$X rays' peaks can be resolved significantly better, which allows to better identify the spectral lines.
The statistics in the peaks are practically unchanged by the cuts, while the background is significantly reduced.
}
\label{fig:qc}   
\end{centering}
\end{figure}

Due to the high rates of the measurements, pileup events contribute to the majority of the events that need to be reduced.
Spurious events that could not be properly reconstructed are also discarded.
The baseline and trigger parameters calculated on the low-frequency waveforms were used for this study.
This study was initially performed using calibration data and then cross-checked with physics data.
The cuts were tested by evaluating the $^{40}$K natural radioactivity line, which is present in both calibration and physics runs, and were studied sequentially in the following order:
\begin{itemize}
    \item[1.-] \textbf{Trigger Number:} traces where only one trigger is registered are accepted.
    \item[2.-] \textbf{Maximum Amplitude Position:} excludes those events whose maximum amplitude occurs after a selected position in the trace.
    \item[3.-] \textbf{Trigger Position:} accepts events occurring within a selected trigger window of the trace. 
    \item[4.-] \textbf{Baseline Exponential Coefficient:} excludes events occurring when the baseline shows a slope.
    \item[5.-] \textbf{Baseline Sigma:} accepts only the events where the baseline fluctuations are below a certain value.
\end{itemize}
The first two criteria help reduce in-trace pileups, which occur when the same waveform contains multiple signals.
The last two target pre-trace pileups, which are signals sitting on the tail of a previous event.
Criterion 3 focuses on selecting events which were properly reconstructed.
The criteria 1 to 3 are common to all detectors.
However, the values of criteria 4 and 5 are evaluated detector by detector, given their different noise conditions.
Fig.~\ref{fig:qc} demonstrates the impact of applying the quality cuts on the $\mu$X-ray lines in the $^{136}$Ba-II data.
In general, 75\,$\%$ to 95\,$\%$ of events survive the cuts in the region of interest -- the spectral peak -- depending on the detector. 
It is clear from the figure that the cuts predominantly remove the background events. 
The energy resolution (discussed in more detail in Sec.~\ref{sec:energy-res}) and the peak-to-background of the $\mu$X rays' peaks is also improved.
\subsubsection{Energy calibration}\label{sec:calibration}
Several calibration runs were done before and after each of the three physics runs to calibrate the spectra. 
The low-energy calibration (up to $\sim$1500\,keV) is done with a $^{152}$Eu source.
To probe the high-energy part (to $\sim$4000\,keV), $\mu$X rays from $^{136}$Ba and the background $\gamma$ ray from $^{208}$Tl were used.

\begin{figure}[t!]
\begin{centering} 
\includegraphics[width=1\columnwidth]{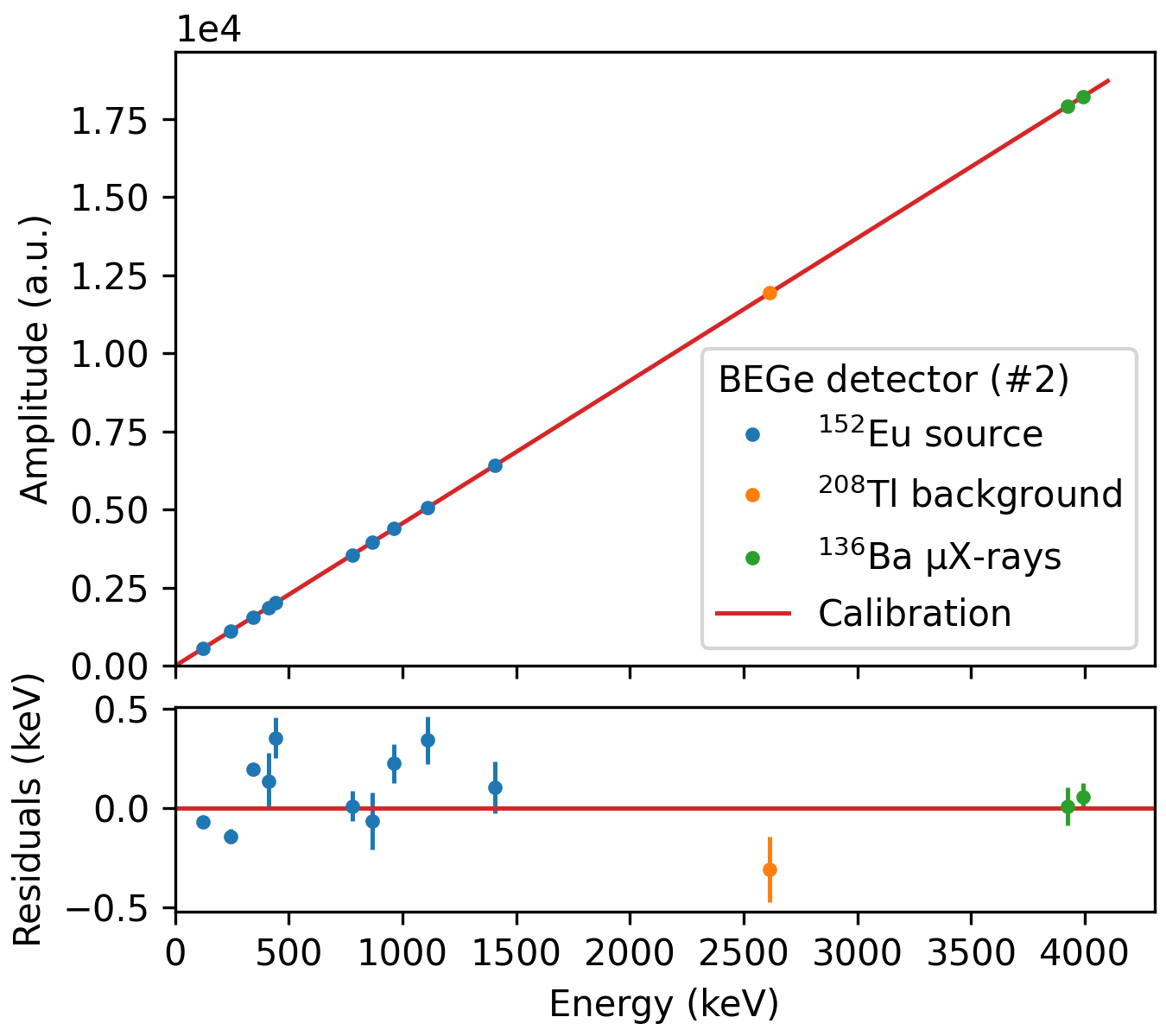}
\caption{
Calibration curve for one of the BEGe detectors ($\#$2) using ALPACA data.
In this case, the calibration curve is a linear function.
}
\label{fig:calibration}   
\end{centering}
\end{figure}

Fig.~\ref{fig:calibration} shows an example calibration curve for one of the HPGe detectors using ALPACA data.
The procedure followed for the MIDAS analysis was the same.
The example in the figure uses a linear calibration function, while for some of the detectors a quadratic calibration is used. 
The choice of the function is made for each of the detectors individually based on the calibration residuals, shown in the bottom panel of the figure.

%% file: 5-performances.tex
\section{Performance}\label{sec:performances}
Some of the main observables are the intensity and position of the spectral lines. 
To extract these quantities, it is necessary to fit first their characteristic shapes.
Although their shape varies from detector to detector, it was found that the peak fit model used by \textsc{Gerda}~\cite{GERDA:2021pcs} suits the data reasonably well, and it is flexible enough to account for the differences among detectors.
It consists of four components: a Gaussian peak, a low-energy (left) tail, and a polynomial and step function to model the background.
Divided into peak and background components, the model is described as follows:
\begin{equation}
     f_{\mathrm{peak}} = n \cdot \left[  (1-\alpha) \cdot f_{\mathrm{gauss}} + \alpha \cdot f_{\mathrm{tail}} \right]   
\end{equation}
where the Gaussian is
\begin{equation}
    f_{\mathrm{gauss}}(E, \mu, \sigma) = \frac{1}{\sqrt{2 \pi \sigma}}\cdot\exp\left[ \frac{-(E - \mu)^2}{2\sigma^2} \right]
\end{equation} 
and the low-energy tail
\begin{equation}
\begin{aligned}
    f_{\mathrm{tail}}(E,\mu, \sigma, \beta) &=  \frac{1}{2\beta} \cdot \exp\left( \frac{E-\mu}{\beta} + \frac{\sigma^2}{2\beta^2}\right) \\
    &\cdot \text{erfc}\left( \frac{E-\mu}{\sqrt{2}\sigma} + \frac{\sigma}{\sqrt{2}\beta}\right)
\end{aligned}
\end{equation}
The amplitude of the peak is given by \textit{n}, and $\alpha$ is the fraction of the tail in the peak. The $\sigma$ value accounts for the resolution of the peak, $\mu$ is the peak position, $\beta$ is the slope of the tail, and \textit{E} stands for energy.
When multiple peaks appear in the fit energy window, they share the same $\sigma$ and $\alpha$ values.
For the background, the equations used are:
\begin{equation}
\begin{aligned}
    f_{\mathrm{bkg}} &=  f_{\mathrm{pol}} + f_{\mathrm{step}}
\end{aligned}
\end{equation}

being the polynomial

\begin{equation}
\begin{aligned}
    f_{\mathrm{pol}}(E, p_{i}) &=  \sum_{i} p_{i} \cdot E^{i}
\end{aligned}
\end{equation}

and the step function

\begin{equation}
\begin{aligned}
    f_{\mathrm{step}}(E, \mu, \sigma, s) &= \frac{s}{2} \cdot \text{erfc}\left(\frac{E - \mu}{\sigma \sqrt{2}}\right)
\end{aligned}
\end{equation} 

The step function is only used when required, with $s$ indicating the size of the step. When multiple peaks are present, the step function is centered at each of the peaks where it is needed. The full mathematical model written to include the possibility of multiple peaks in the fit window is:

\begin{equation}
\begin{aligned}
    \label{eq:model}
    f_{\mathrm{model}}(\mu_{i}, n_{i}, p_{i}, s_{i}, \ldots) &= \sum_{i} \left( f_{\mathrm{peak}}(\mu_{i}, n_{i},\ldots) \right. \\
    &\quad \left.+ f_{\mathrm{bkg}}(\mu_{i}, p_{i}, s_{i},\ldots) \right)   
\end{aligned}
\end{equation}

Here, the ellipsis indicates all parameters that are common within the fit energy window, such as the $\sigma$ and $\alpha$ values.

In cases where the peak's natural width is significant, e.g. $\mu$X rays, the Gaussian kernel was replaced with a Voigt profile~\cite{ROOT}, which convolutes the Gaussian with a Lorentzian function whose $lg$ parameter is fixed to the natural width calculated with the Mudirac code~\cite{Sturniolo:2020wsq}. Below is the mathematical model of the Lorentzian function and the Voigt profile.

\begin{equation}
\begin{aligned}
    f_{\mathrm{lorent.}}(lg, E) &= \frac{lg}{\pi \cdot (E^2 + lg^2)}
\end{aligned}
\end{equation} 

\begin{equation}
\begin{aligned}
    f_{\mathrm{voigt}}(E ; \sigma, lg) &= \int_{-\infty}^{\infty} f_{\mathrm{gauss}}(E'; \sigma) f_{\mathrm{lorent.}}(E-E'; lg) \,dE'
\end{aligned}
\end{equation}

\begin{figure}[t!]
\begin{centering} 
\includegraphics[width=1\columnwidth]{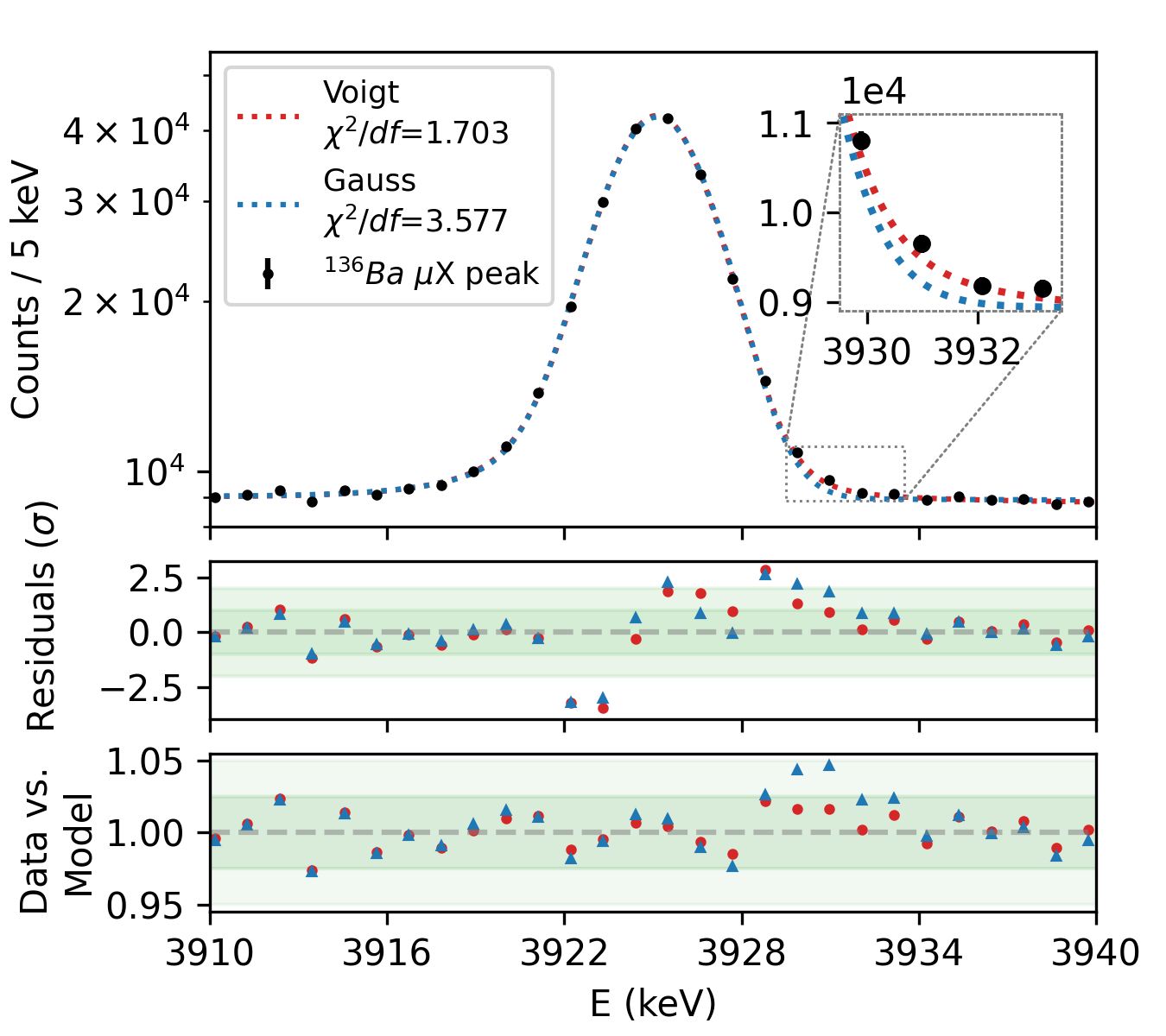}
\caption{
$^{136}$Ba $\mu$X ray at 3925.2\,keV for detector $\#$4 (COAX). 
The peak is fitted both with the modified model using a Voigt profile (red) and with the model using a Gaussian (blue).
The Voigt profile provides a better fit to the right tail of the $\mu$X ray peak.
Another indication that validates the use of a Voigt profile is given by the $\chi^2/df$, which in this case goes from 3.577 to 1.703 when including the Voigt profile.
This value varies between different peaks and detectors, but the improvement in $\chi^2/df$ is usually of about one unit.
}
\label{fig:voigt}   
\end{centering}
\end{figure}

An example can be seen in Fig.~\ref{fig:voigt}, showing a better fit to the non-Gaussian right tail. 
The lifetime-related broadening of the left tail is fitted adequately by both the Voigt and \textsc{Gerda} functions, because the latter contains a dedicated left-tail component to describe incomplete charge collection in HPGe detectors.
\subsection{Energy resolution}\label{sec:energy-res}
\begin{figure}[t!]
\begin{centering} 
\includegraphics[width=1\columnwidth]{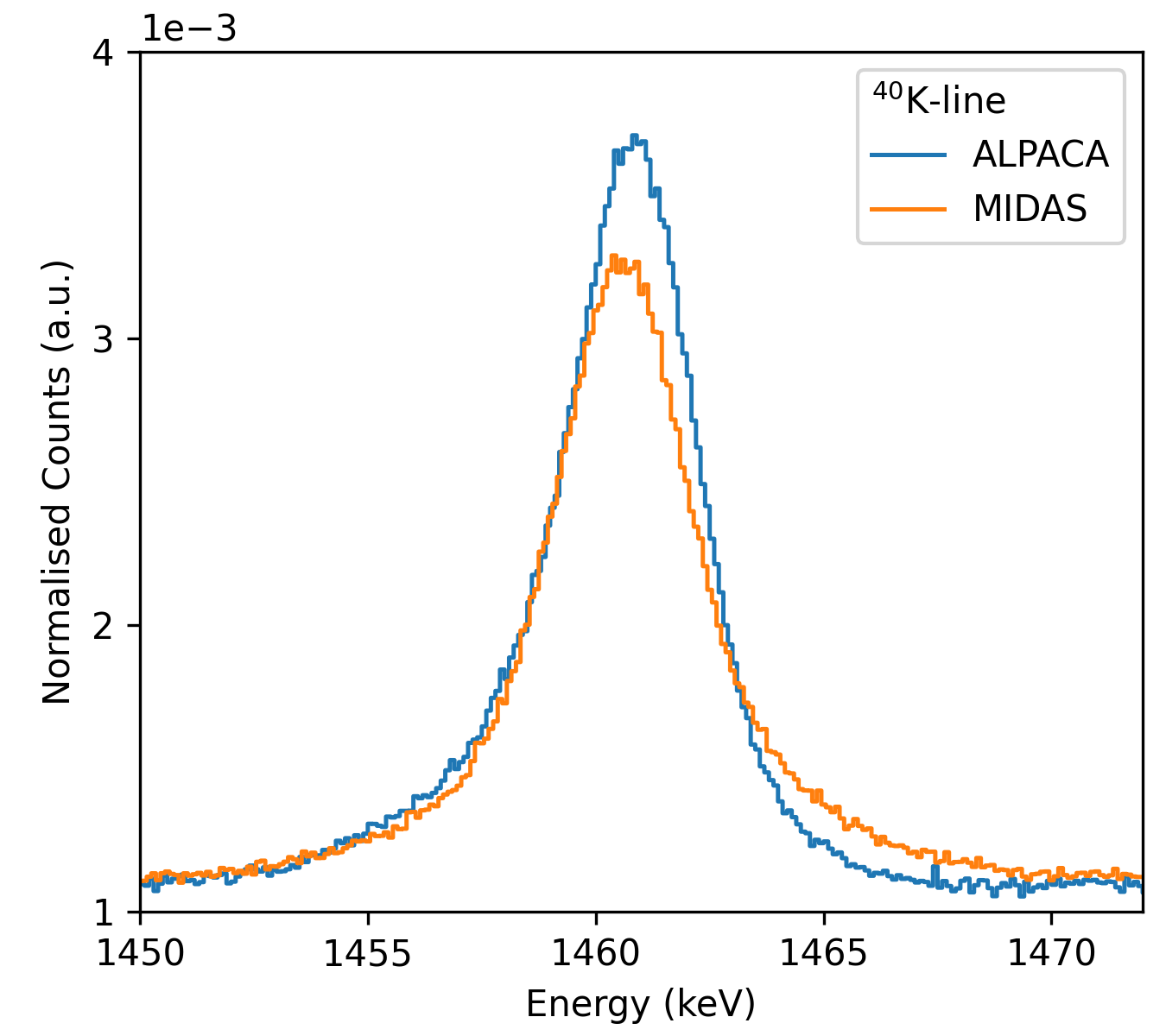}
\caption{
$^{40}$K background peak for a COAX detector ($\#$4) using the $^{136}$Ba data. 
Given the statistical differences between ALPACA and MIDAS, the amplitude of the peak is normalised to be able to do a one-to-one comparison of the peak shape between the two DAQ systems.
The offline DSP from ALPACA enabled a better energy resolution in comparison to MIDAS.}
\label{fig:energy-res}   
\end{centering}
\end{figure}
\begin{table}[b!]
    \centering
    \caption{Energy resolution for detector $\#$2 (BEGe) and $\#$4 (n-type COAX) using the $^{40}$K-line in the $^{136}$Ba and $^{76}$Se data. 
    The resolution of the BEGe detector is significantly better than the COAX's resolution. 
    The slightly better resolution obtained for ALPACA is attributed to the offline energy reconstruction process.
}
    \begin{tabular}{  m{3.2em} | m{1.35cm} m{1.35cm} |  m{1.35cm} m{1.35cm}   }
        \multicolumn{1}{c}{} & \multicolumn{4}{c}{$^{40}$K-line FWHM (keV)}\\     
        \hline
        \multicolumn{1}{c}{}& \multicolumn{2}{c}{Detector 2 (BEGe)}  &  \multicolumn{2}{c}{Detector 4 (COAX)}\\
        \hline
        Runs & MIDAS & ALPACA &MIDAS & ALPACA \\
        \hline
        $^{136}$Ba& 2.35$\pm$0.02 & 2.31$\pm$0.02 & 4.12$\pm$0.02 & 3.62$\pm$0.01\\
        $^{76}$Se& 2.29$\pm$0.02 & 2.15$\pm$0.03 & 3.87$\pm$0.02 & 3.40$\pm$0.02\\
    \end{tabular}
    \label{tab:resolution}
\end{table}
The ALPACA data used in this section have passed the data quality cuts described in Sec.~\ref{sec:qcs}.
The energy resolution for the HPGe detectors was studied using the $^{40}$K background peak, present in calibration and physics runs.
In Table~\ref{tab:resolution}, the FWHM values for two detector types are presented and calculated using all the data from the $^{136}$Ba and $^{76}$Se runs.
In this case, the FWHM was extracted after fitting the data with the aforementioned model in Eq.~\ref{eq:model}.
The fit isolates the peak (Gaussian and tail) from the background components. 
Then, the FWHM is calculated using the peak part only.
The values extracted for BEGe and COAX detectors are compared using ALPACA and MIDAS data.
Fig.~\ref{fig:energy-res} shows the $^{40}$K peak for the COAX detector, demonstrating that the energy resolution is better with ALPACA than MIDAS.
This is attributed to the offline energy reconstruction, that was optimised precisely to improve this parameter, and to the application of quality cuts. 
\subsection{Time resolution}
\begin{figure}[t!]
\begin{centering} 
\includegraphics[width=1\columnwidth]{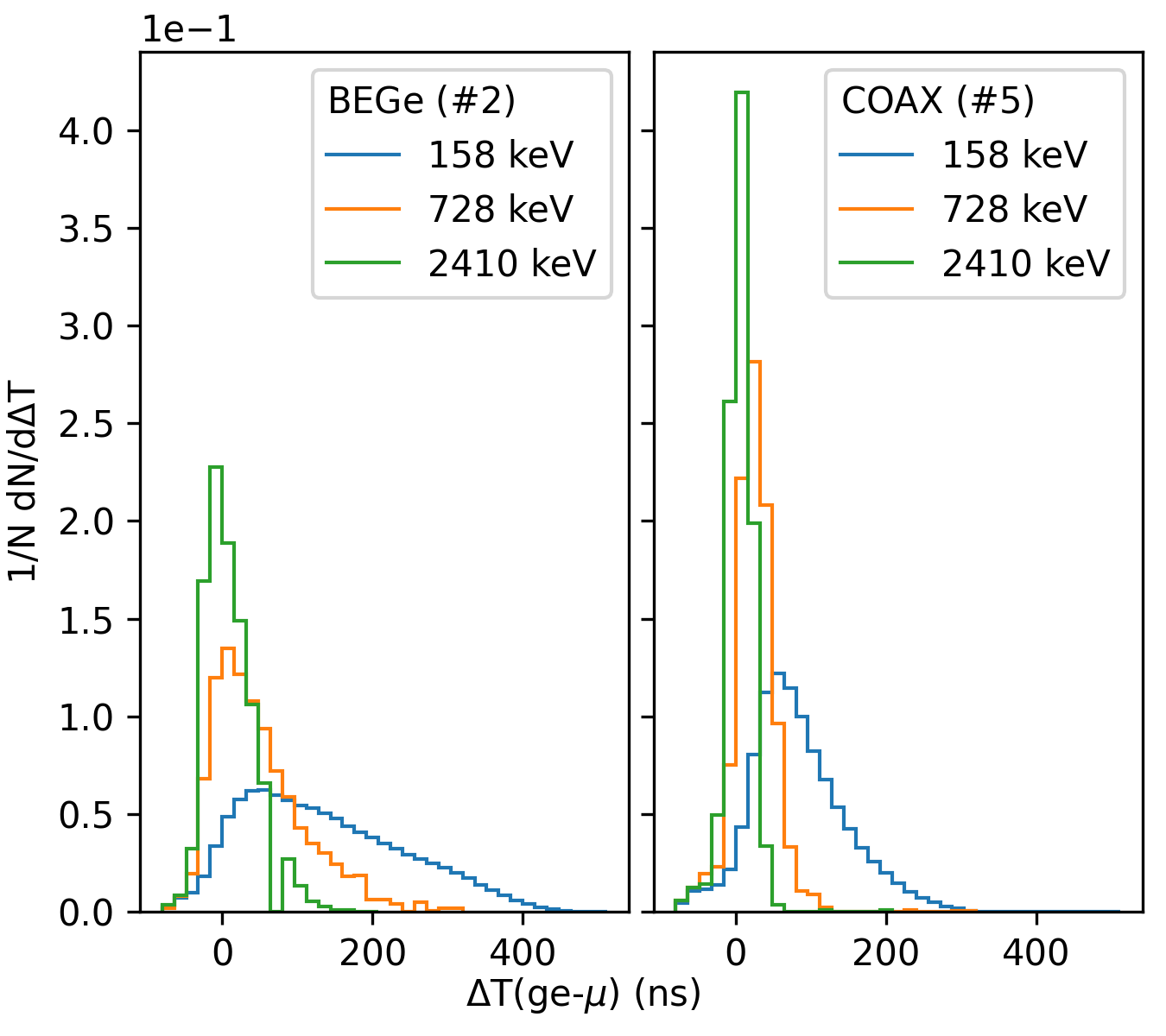}
\caption{
Normalised count rates of $\mu$X rays plotted against the time interval between the signal in the HPGe detector and the muon stop.
The distribution is expected to be centred around zero since the $\mu$X rays cascade emission characteristic of the muonic atom is of $O(ps)$, which far exceeds the precision of the system $O(ns)$.
It is observed, however, that the lower the energy, the skewer the distribution towards longer times, worsening the time resolution.
This effect is less prominent in COAX compared to BEGe detectors (see text).
}
\label{fig:time-resolution}   
\end{centering}
\end{figure}
Proper reconstruction of the muon lifetime requires good time resolution.
The time resolution depends on the ability to precisely determine the trigger position of a germanium event in relation to the correlated event from the counters.

The DAQ systems have a sampling period of a few nanoseconds, making them unable to resolve events in the picosecond regime.
Therefore, it can be considered that a $\mu$X-ray event is detected simultaneously in the HPGe detector and in the counters.
Given a finite time resolution in the HPGe detectors, a Gaussian-like distribution centred in zero is expected when studying these events projected in time.

However, an energy-dependent relation is observed, as shown in Fig.~\ref{fig:time-resolution}, quantified by the RMS, reaching from 20 ns at high energies up to almost 10 ns at low energies. 
Lower energy pulses often have longer rise times, which worsens the time reconstruction, and thus, the time resolution also worsens towards the low energies.

COAX detectors typically achieve better performance than the BEGe's due to the BEGe's intrinsic longer rise times, which originate from the longer charge-carrier drift time.
Also, BEGes have a broader range of rise times than other HPGe detectors.
Nonetheless, the effect is seen in all detectors.
\subsection{Relative detection efficiency}\label{sec:efficiency}
The relative efficiency is obtained by extracting the $\gamma$ rays' intensities from $^{152}$Eu and $^{\textrm{nat}}$Pb and dividing each line by its branching ratio.
$^{\textrm{nat}}$Pb is advantageous to probe the high-energy end of the spectrum because the K-series $\mu$X rays reach
$\sim$ 6 MeV.
Having other $\mu$X-ray lines at lower energies made it possible to compare the efficiency curve with the one obtained with the $^{152}$Eu source alone.

For this comparison, a combined fit was performed using the values from both sources. 
Each data set uses an individual normalisation -- which accounts for the source intensity -- but shares common shape parameters.
The data from different calibration periods are also combined and compared.
The resulting efficiency curves show consistency among different periods and sources.

\begin{figure}[t!]
\begin{centering} 
\includegraphics[width=1\columnwidth]{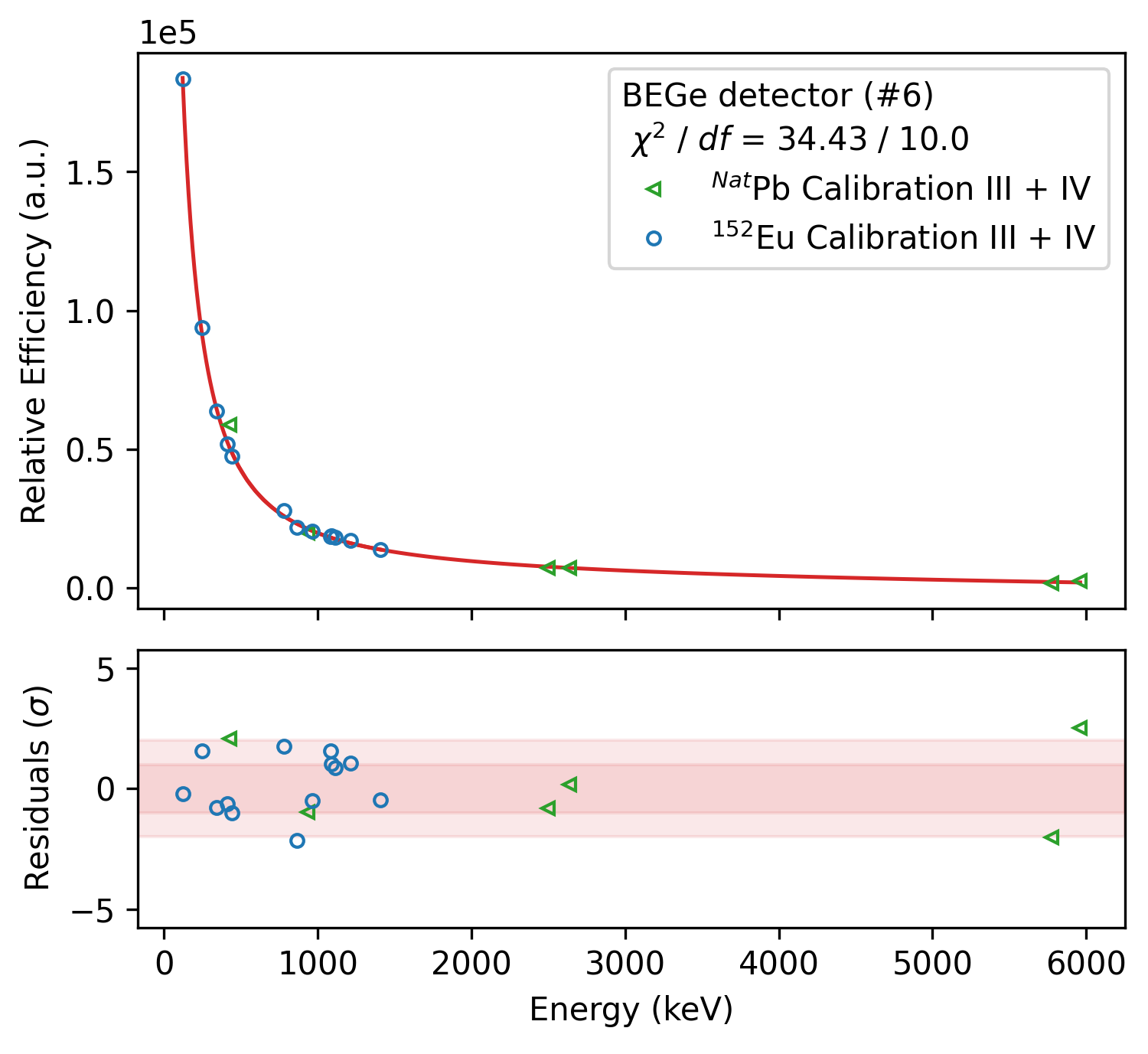}
\caption{
Combined relative efficiency fit that includes $^{152}$Eu and $^{\textrm{nat}}$Pb data from two different calibration periods (before and after the $^{136}$Ba-II run)
The chosen polynomial function provided the best fit in this case based on the residuals and the $\chi^{2}/df$ value. 
}
\label{fig:efficiency}   
\end{centering}
\end{figure}

Empirical polynomial functions are used to fit the curves.
Fig.~\ref{fig:efficiency} shows the efficiency curve obtained for detector $\#$6 (BEGe) produced with ALPACA data using the following model~\cite{efficiency}:

\begin{equation}
\label{eq:poly}
\epsilon(E) = \frac{1}{E}\cdot\sum_{i} C_{i}\ln(E)^{i}\
\end{equation} 
The functional form of this empirical description is chosen based on the $\chi^{2}/df$ value, and under study detector by detector.

\subsection{Time stability}
\begin{figure*}[t!]
    \centering
    \includegraphics[width=\textwidth]{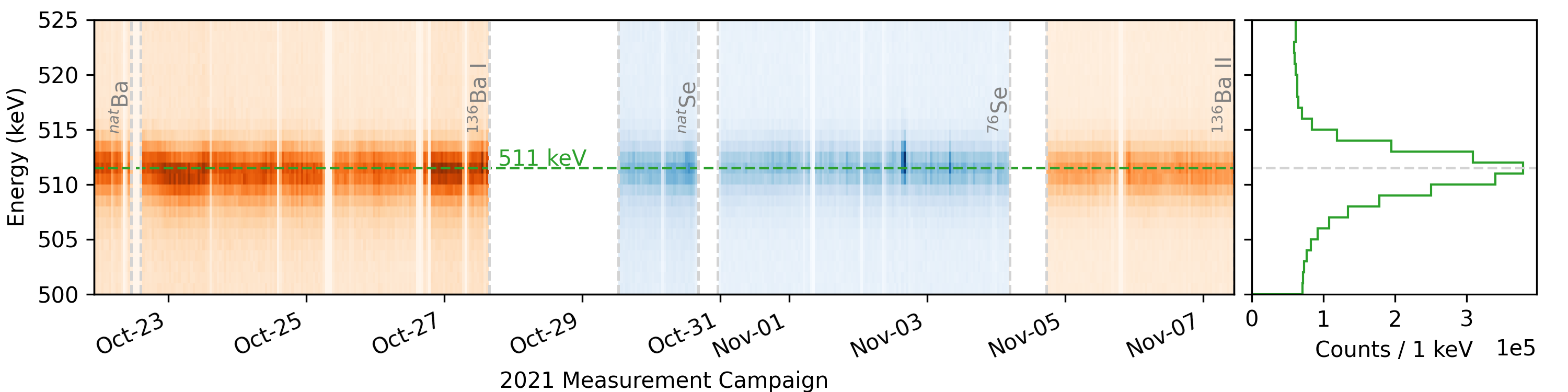}
    \caption{
\textbf{Left:} density plot showing the position of the 511 keV $\gamma$-ray line during the 2021 measurement campaign for detector $\#$4 (n-type COAX).
The line position remains stable over time for the different data-taken periods.
\textbf{Right:} 511\,keV peak using all data shown in the left plot. 
}
    \label{fig:time_stab}
\end{figure*}
It was studied whether the energy position of the spectral lines was drifting over the duration of the measurement campaign.
Possible reasons for the drift are temperature and humidity variations, vibrations produced after the refilling of the detectors, and electronic issues, among others.
If drifting is observed, one needs to understand the scale of the drifting and whether the variations follow any pattern. 
This allows one to apply an offline correction, preserving the energy resolution.
However, if the drift is small --within the uncertainty on the energy position-- there is no need to correct it.

By studying the position of several spectral lines over the beam time, it was observed that it was fairly stable.
Fig.~\ref{fig:time_stab} shows the 511\,keV line over the whole measurement campaign.
Although there are spectral lines with more statistics, the 511\,keV line is common to both the $^{136/\textrm{nat}}$Ba and $^{76/\textrm{nat}}$Se data enabling continuity.
The $\mu$X rays at the high-energy end were also studied, and no significant drifting in the energy position was observed.

It is concluded that the data-taking went under sufficiently stable conditions that did not significantly affect the energy position of the spectral lines.
\subsection{DAQ Livetime}\label{sec:livetime}
\begin{figure*}[t!]
    \centering
    \includegraphics[width=\textwidth]{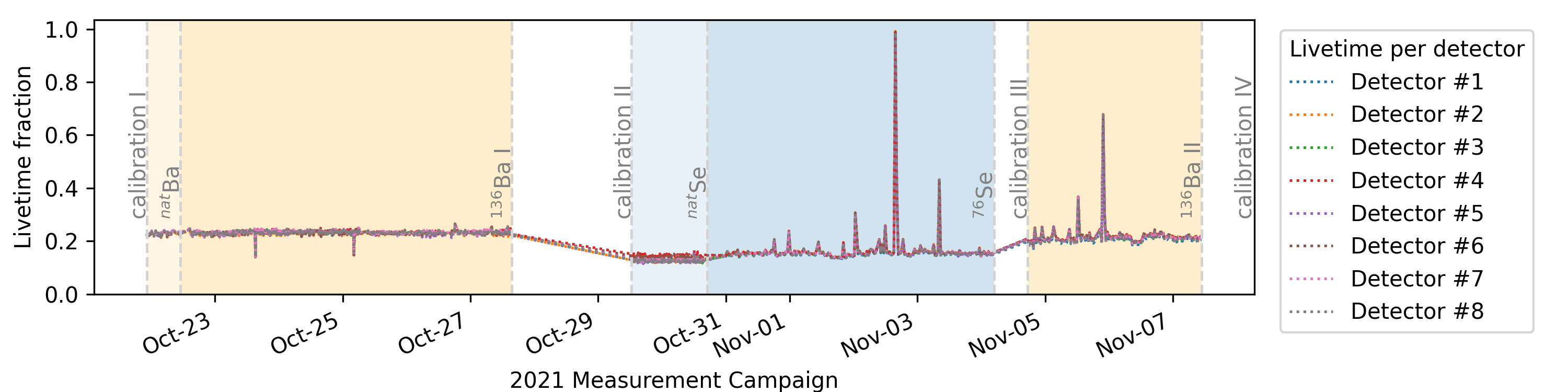}
    \caption{
    Time-resolved combined livetime fraction for all germanium channels during the entire data-taking period employing enriched targets.
    The combined livetime fraction measures the performance of ALPACA, encompassing the livetime of single germanium channels and the availability of all PMT channels.
    Results on either side are obtained with different analysis procedures, dependent or independent of the test pulse (see text).
    }
    \label{fig:deadtime}
\end{figure*}
Unlike MIDAS, ALPACA is a new system, and its deadtime performance is not yet understood. 
To quantify it, three characteristics were evaluated: 
\begin{itemize}
    \item[1.-] The loss in recorded signals due to the bank swap feature (explained in Sec.~\ref{sec:daq}).
    A problem arises in case of a high data rate, exceeding the speed of the ethernet interface measured to be around 600 Mbit/s.
    In this case, some channel buffers are filled before the data transfer from the passive bank finishes. 
    Thus, these channels show deadtime until the readout finishes and the next bank swap occurs.
    This effect differs between detectors since they fill their memory buffers at different speeds. 
    This is due large difference in individual trigger rates, which originate from characteristic detection efficiencies and trigger thresholds.
    \item[2.-]Another source of deadtime arises when a new file is being created.
    A gap, with no pulses being recorded for several seconds, spans from the end of one file to the beginning of the next one.
    Presumably, it is caused by the server's operating system finishing writing the first file to the hard drive, thus stalling the program.
    All channels are affected equally by this effect since the program issues the command to resume data-taking only after a new file is created.
    \item[3.-]In-trace pileups lead to an additional source of deadtime.
    They are caused by the FADC ignoring any potential trigger in a channel while still processing a previous signal.
    The deadtime induced by each processed signal is defined by the "event window'', whose size depends on the maximum time duration of online processing and/or sampling.
    For ALPACA, the low-frequency traces define the event windows, which have a duration of 19.2 $\mu$s for germanium channels and 7.04 $\mu$s for PMT channels. 
\end{itemize}
A dedicated algorithm to calculate the livetime reduction due to the two types of gaps for every detector channel was developed.
It investigates differences in timestamps of consecutive events and counts all gaps exceeding 100 ms as deadtime, excluding those larger than 60 s as those are caused by human interaction.
The resulting livetime after considering this effect is $\sim$77$\%$ for $^{76}$Se and of $\sim$88$\%$ for $^{136}$Ba runs.

The livetime reduction caused by pileups is calculated by studying the time difference of consecutive events, which follows a Poissonian distribution.
Extrapolating the statistical distribution towards the lower end ($\Delta$t$=$0) makes it possible to compute the difference with the actual distribution and account for the fraction of missed events.

Thus, livetime loss observed in individual channels results separately from the effects of the gaps and deadtimes introduced by pileups.
Effective livetimes of single channels are obtained by combining the livetime fractions derived from both effects.
This way, after including the pileups contribution, the livetime is reduced to $\sim$73$\%$ for $^{76}$Se and to $\sim$81$\%$ for $^{136}$Ba.

An independent analysis was done to evaluate the livetime.
For this separate analysis, the measured frequency of a high-precision test pulse sent to all detectors simultaneously is used.
By evaluating the ratio between injected and measured test pulses, it is possible to calculate the fraction of non-recorded events \footnote{This method is only available for some data-taking periods.}.
When comparing the results from the two independent analyses, practically the same livetime fraction was found, thus validating these results.

The term ``combined livetime'' further includes the PMT channels' contribution.
Apart from recording the waveform from a triggering HPGe detector, ALPACA needs to simultaneously record waveforms from all four PMTs. 
The combined livetime is reduced when the system misses at least one of the PMT waveforms accompanying the HPGe's trigger.
This effect also equalises the livetime fraction across all detectors, given that each of the PMT channels' buffers is saturated at the same rate, regardless of which HPGe triggered the event.
During the 2021 measurement campaign, which was the first time ALPACA was deployed, it turned out that this contribution was dominating the livetime loss, decreasing the average livetime fraction from $\sim$80\,$\%$ to $\sim$20\,$\%$ ($\sim$16$\%$ for $^{76}$Se and to $\sim$23$\%$ for $^{136}$Ba runs). 
In the future, the plan is to mitigate this effect by reducing the length of the low-frequency PMT traces from 7 $\mu$s to 4-5 $\mu$s, which should increase the time until the PMT channels' buffer saturates.
If not enough, the low-frequency traces can be dispensed altogether, since the main purpose of the PMT signal is the triggering information.
Other ideas include further compression of waveforms and developments in the ALPACA software that help mitigate the bank-swapping deadtime.
Fig.~\ref{fig:deadtime} shows the combined livetime extracted for the physics data. 
The livetime is mostly constant with the exception of short-term spikes that are attributed to instabilities in the muon-beam.
\subsection{Trigger Rates}\label{sec:trigger-rates}
\begin{figure*}[t]
  \centering
  \begin{subfigure}[t]{0.49\textwidth} 
    \includegraphics[height=8cm, width=\textwidth]{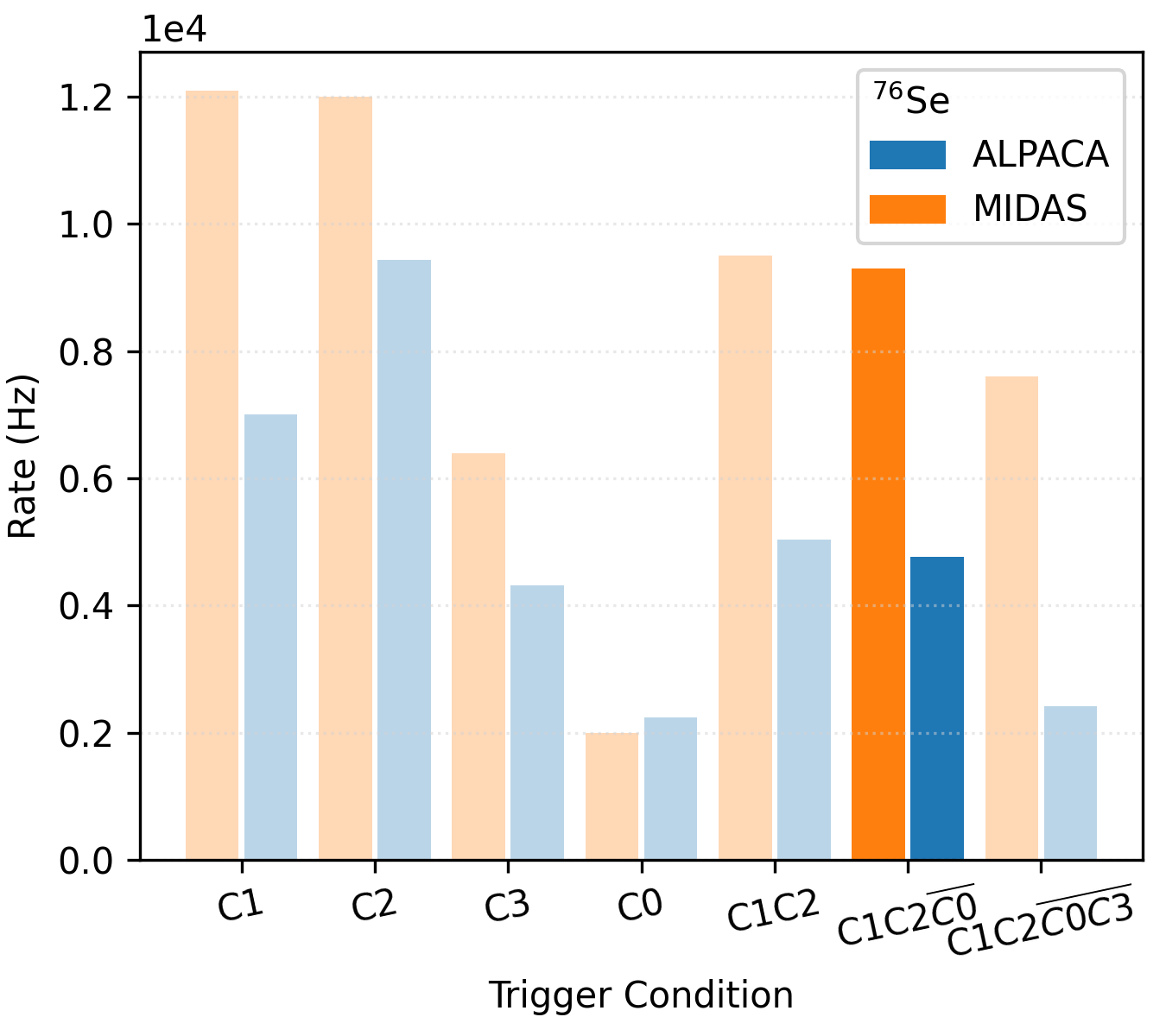} 
    \caption{
      Event rates for the different trigger conditions studied for one of the HPGe-detectors using $^{76}$Se data.
      The originally considered trigger condition (the rightmost entry) resulted in a severe reduction of the event rate. 
      The trigger condition ultimately chosen for data analysis is highlighted.
    }
    \label{fig:trigger-scheme} 
  \end{subfigure}
  \hfill
  \begin{subfigure}[t]{0.49\textwidth} 
    \includegraphics[height=8cm, width=\textwidth]{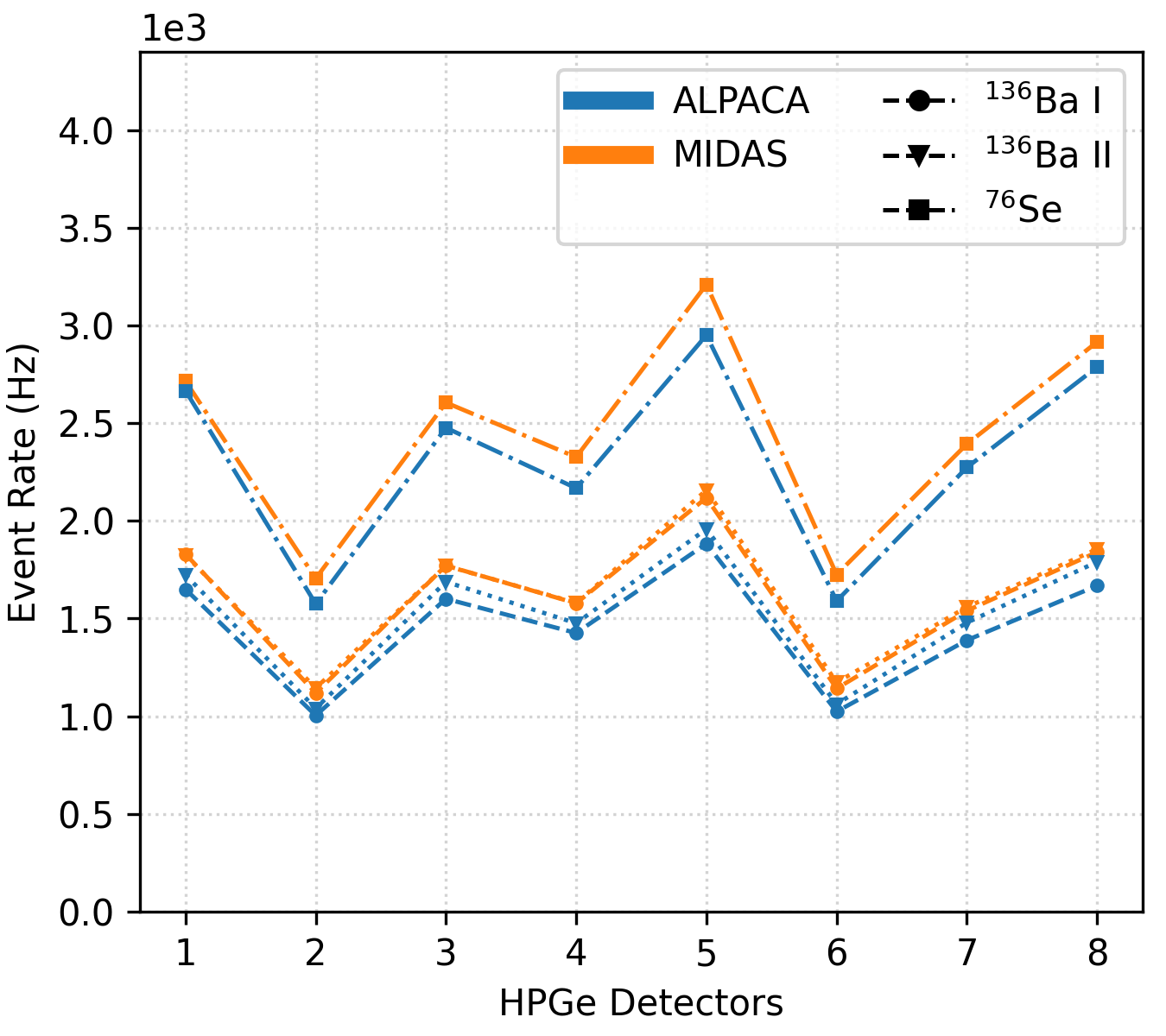} 
    \caption{
      Event rates from the $^{136}$Ba and $^{76}$Se runs for ALPACA and MIDAS.
      The event rates from $^{76}$Se are slightly higher than for $^{136}$Ba, possibly due to the higher probability for muons to undergo OMC in $^{76}$Se, as compared with $^{136}$Ba.
      It is also seen that the rate is the highest for detector $\#$5, a n-type COAX detector with the highest efficiency.
      Correspondingly, the two BEGe detectors -- $\#$2 and $\#$6 -- present the lowest rates from the array.
    }
    \label{fig:event-rates} 
  \end{subfigure}
  \caption{(a)~The rate of events for a HPGe detector when applying different trigger conditions. (b)~The total event rates per detector observed during the measurement campaign.
  }
  \label{fig:rates}
\end{figure*}
\begin{figure*}[t!]
    \begin{centering} 
    \includegraphics[width=\textwidth]{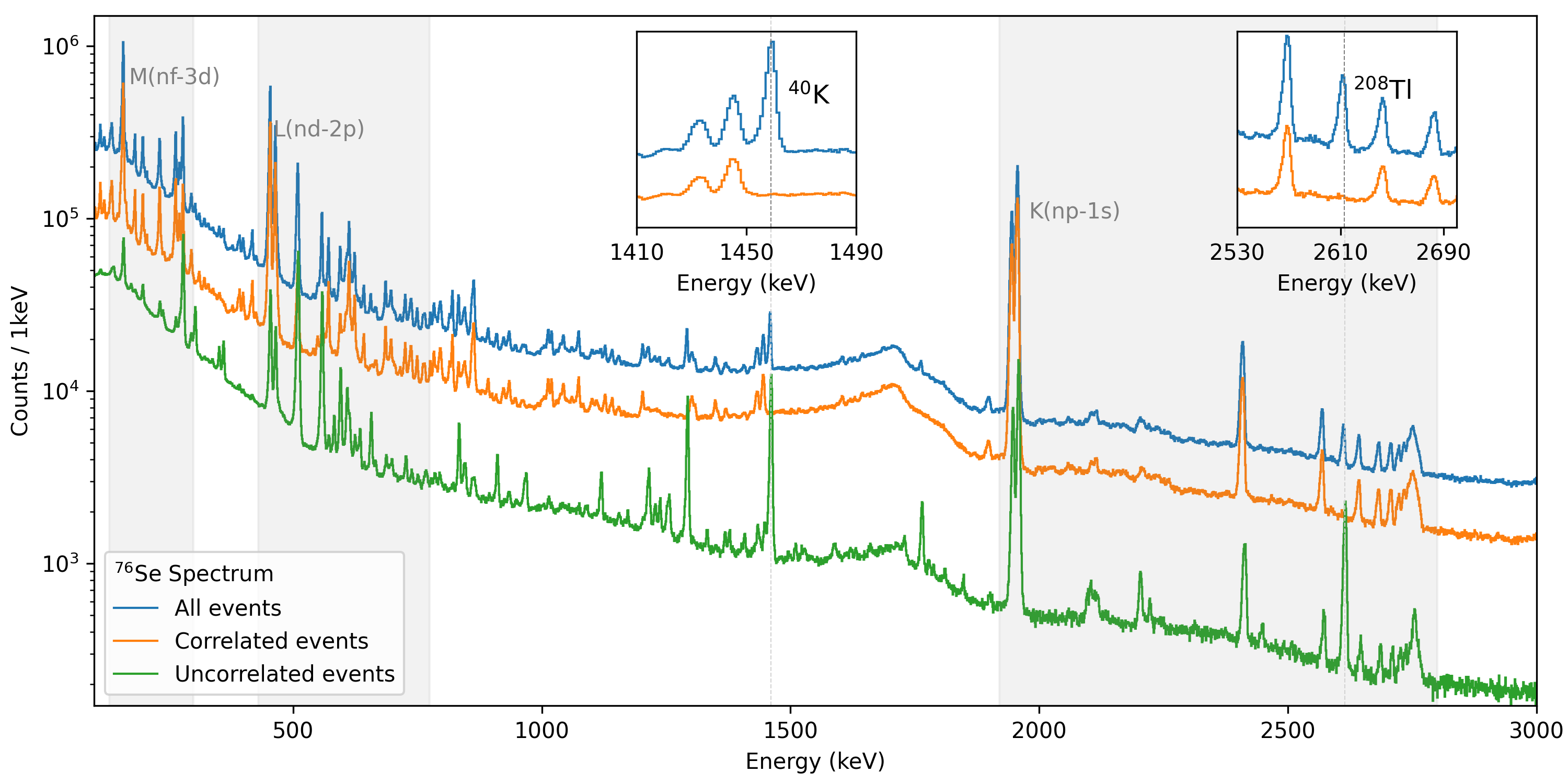}
    \caption{
    $^{76}$Se spectra before (blue) and after (orange) applying the trigger condition to tag coincident muons for detector $\#$5 (COAX n-type with an estimated 95$\%$ detection efficiency).
    The uncorrelated spectrum (green) is built by subtracting the correlated spectrum from the total spectrum.
    The two insets highlight how the background lines, e.g., $^{40}$K and $^{208}$Tl, disappear in the \textit{correlated} spectrum, as expected when applying a coincidence criterium between the measured $\gamma$ rays and the muon signals from the counters.
    }
    \label{fig:corrUncorr}   
    \end{centering}
\end{figure*}
The OMC trigger condition shown in Eq.~\ref{eq:trigger_omc} produces the cleanest sample of stopped muons, but it turned out to remove a significant amount of valid events.
Thus, it was necessary to evaluate the trigger rates of the individual counters to understand if a less stringent trigger criterion could result in higher acceptance rates.
For completeness, the trigger rates in each individual counter in coincidence with HPGe detectors were also evaluated: C$_{i}$-only (i=0,1,..,3).
Additionally, the conditions C$_{1}\wedge$C$_{2}$ and $\overline{\textrm C}_0\wedge$C$_{1}\wedge$C$_{2}$ were investigated as well.

Whenever a coincidence between a HPGe detector and a counter is selected, the time coincidence window extends from -200 to 1000 ns and the energy of the HPGe detector event is requested to be between 100-4500~keV. 
No additional quality cuts were applied to consider the totality of the events and make the results comparable between MIDAS and ALPACA. 
The coincidence/anticoincidence between two counters is defined in the time window between -100 and 100 ns.

These conditions are studied using the two $^{136}$Ba runs and the $^{76}$Se one. 
Fig.~\ref{fig:trigger-scheme} shows the different rates for one detector after applying each listed trigger condition using $^{76}$Se data.
The extracted trigger rates for MIDAS and ALPACA are on the order of a few kHz.
The ALPACA rates are lower due to the deadtime discussed in the previous section.
It is not clear why the C$_0$-only condition provides a higher rate in ALPACA in comparison with MIDAS, but it could be due to the fact that this is an anti-coincidence counter, and hence lower anti-coincidence rates result in higher total-event rates.
This should, however, also apply to the C$_3$-only case, where no higher rate is observed in ALPACA.
The C$_3$ energy spectrum is currently not fully understood, and further studies that clarify its use are ongoing.
Using the initial trigger criterion is found to reduce the amount of accepted valid events by 40$\%$ when compared to the case without using the C$_3$ counter's spectrum.
Therefore, it was opted to use \(\overline{\textrm C}_0 \wedge {\textrm C}_1 \wedge {\textrm C}_2\) condition instead. 
Fig.~\ref{fig:corrUncorr} shows how the chosen trigger condition is sufficient to enhance the signal-to-background in the correlated spectrum.

The event rate per HPGe detector for each of the physics runs is shown in Fig.~\ref{fig:event-rates}. 
The ALPACA rate includes a correction for the deadtime discussed in Sec.~\ref{sec:livetime}, which brings the ALPACA and MIDAS rates in reasonable agreement. 
The residual discrepancy could be due to the differences in the way the rates are extracted for the two systems but is not considered large enough to warrant further study.
The individual detectors' rates depend strongly on the type of detector and its efficiency. 
For example, the BEGe detectors ($\#$2 and $\#$6), whose estimated relative size is 38$\%$, show the lowest rates.
In contrast, detector $\#$5, an n-type COAX with an estimated relative size of 95$\%$, presents the highest rates of all.
\subsection{Random Coincidences}\label{sec:random-coinc}
\begin{figure}[t!]
\begin{centering} 
\includegraphics[width=1\columnwidth]{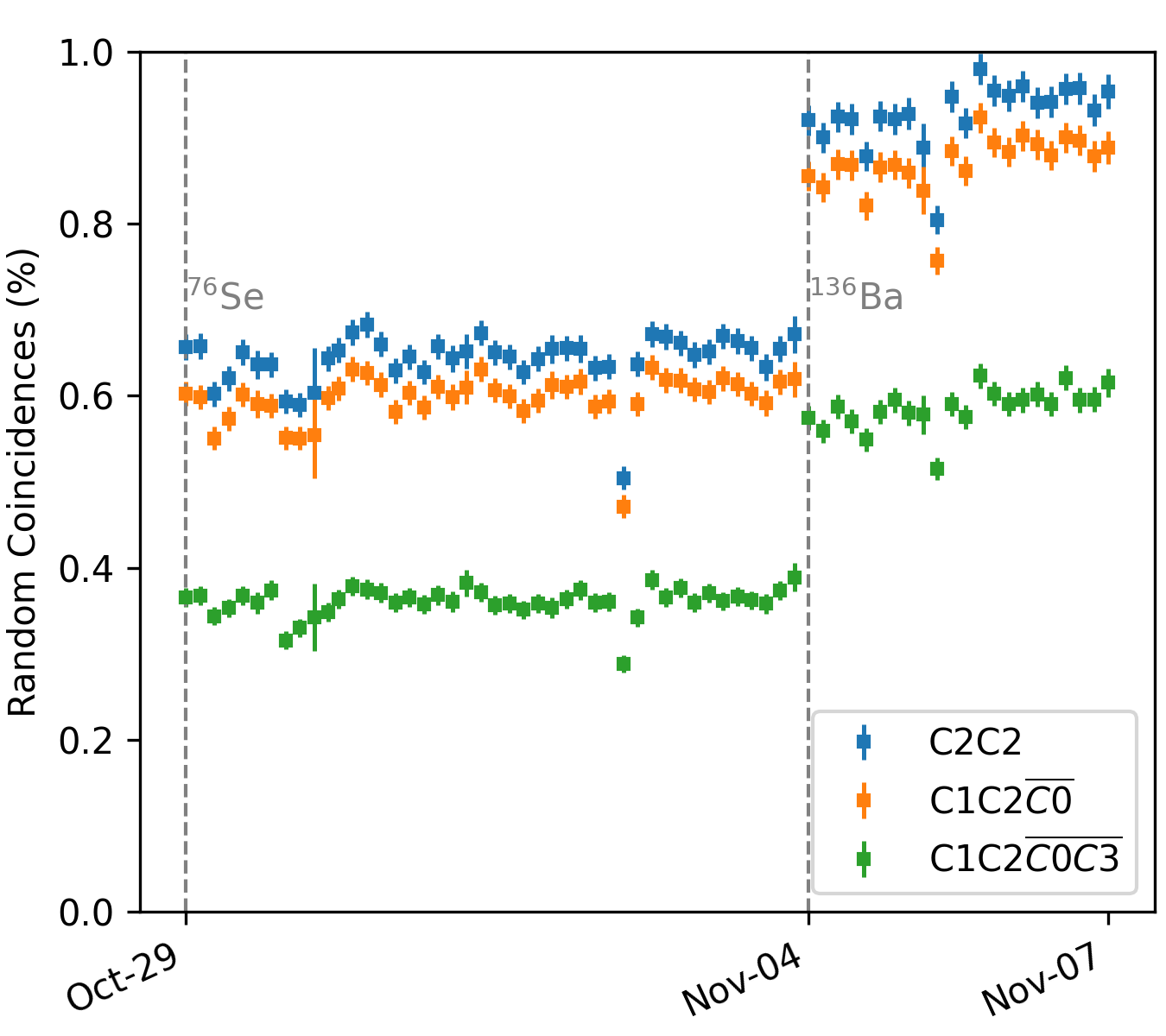}
\caption{
Fraction of random coincidences in the physics data for three different trigger conditions.
Although the default trigger condition $\overline{\textrm C}_0 \wedge {\textrm C}_1 \wedge {\textrm C}_2 \wedge \overline{\textrm C}_3$ results in a smaller fraction, the criterion $\overline{\textrm C}_0 \wedge {\textrm C}_1 \wedge {\textrm C}_2$ to flag valid events was chosen for the reasons explained in Sec~\ref{sec:trigger-rates}.
The difference between events selected using ${\textrm C}_1 \wedge {\textrm C}_2$ and $\overline{\textrm C}_0 \wedge {\textrm C}_1 \wedge {\textrm C}_2$ conditions are minimal when extracting the random coincidences' fraction.
}
\label{fig:random-coinc}   
\end{centering}
\end{figure}
The last presented parameter is the fraction of random coincidences, which are events that pass the trigger condition but are not muon-related.
To estimate it, periodic (30~Hz) test pulses were issued using a high-precision pulser unit during the $^{76}$Se and $^{136}$Ba-II runs, and the fraction of the pulses that passed the trigger condition was recorded.
Fig.~\ref{fig:random-coinc} shows that the random coincidence fraction is less than 1\,$\%$ for the three considered trigger conditions during the whole duration of the runs.
For the original trigger condition (Eq.~\ref{eq:trigger_omc}), the fraction is twice as low as for the other two conditions. 
This effect must be related to the C$_3$ counter's spectrum, which is not yet fully understood.

The fact that the random coincidence fraction is smaller during the $^{76}$Se run is likely due to this run's higher physics event rate. 
The accordingly higher fraction of muon-related correlated coincidences then dilutes the contribution of the random ones.

%% file: 6-conclusions.tex
\section{Summary and outlook}\label{sec:conclusions}
\begin{figure*}[t]
  \centering
  \begin{subfigure}[t]{0.49\textwidth} 
    \includegraphics[height=8cm, width=\textwidth]{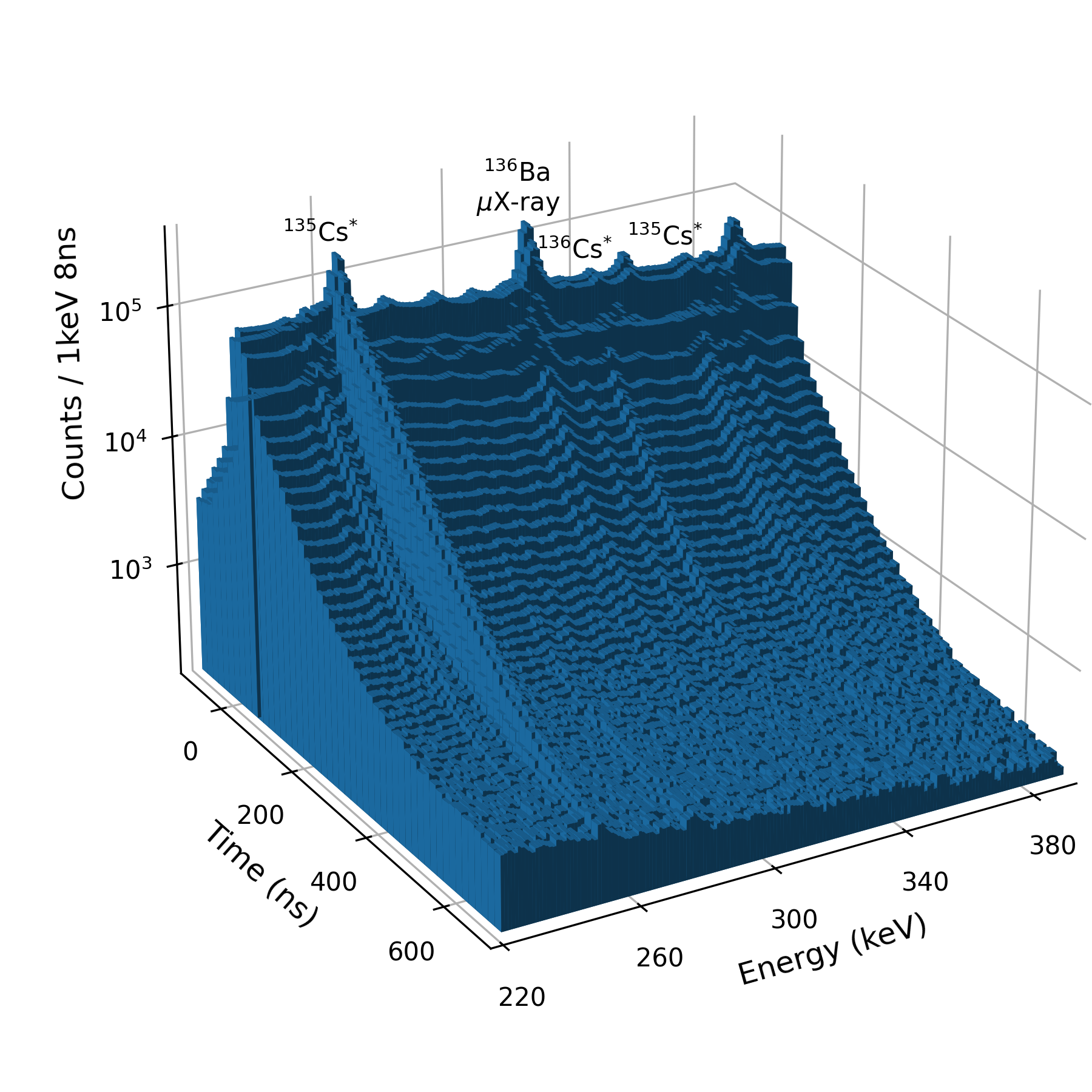} 
    \caption{
    2D histogram showing a small energy window from the $^{136}$Ba correlated spectrum. 
    The $\gamma$ rays decay over hundreds of nanoseconds.
    The $\mu$X-rays decay within tens of nanoseconds.
    Some $\gamma$ rays are more intense than others, which allows the extraction of the partial capture rates based on these intensity ratios.
    }
    \label{fig:2d-hist_a} 
  \end{subfigure}
  \hfill
  \begin{subfigure}[t]{0.49\textwidth} 
    \includegraphics[height=8cm, width=\textwidth]{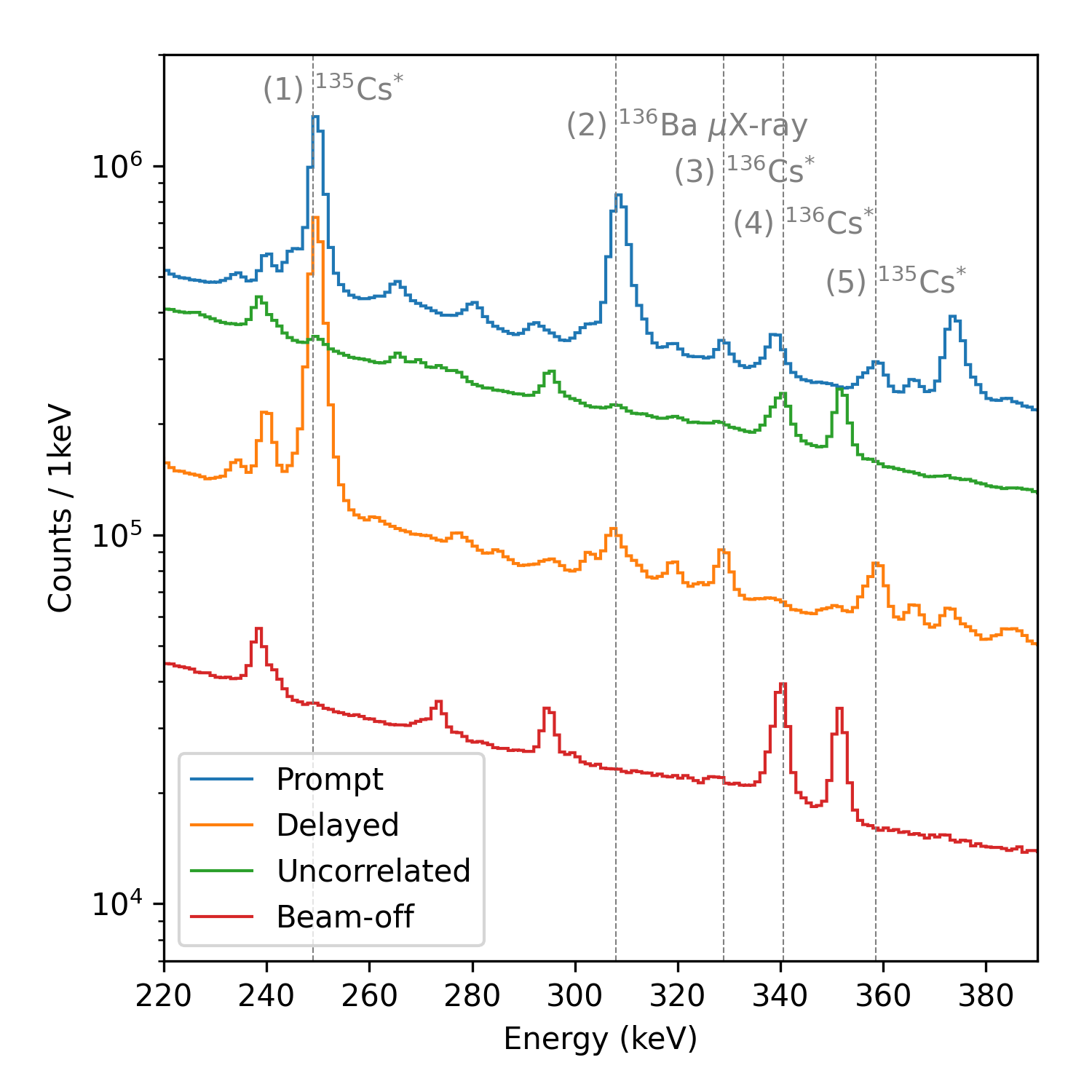} 
    \caption{
    Several spectra based on different cuts. 
    Shown are the \textit{correlated-prompt} and \textit{-delayed} spectra and, in direct comparison, the \textit{uncorrelated} spectrum, which excludes all events from the correlated one.
    One can appreciate how in the prompt and delayed spectra, $\mu$X-rays and $\gamma$ rays are separated. 
    See, for example, the evolution of lines (1), (3) and (5) in comparison with (2), a $\mu$X-ray from the $^{136}$Ba target.
    (4) corresponds to a $\gamma$-ray line with a lifetime longer than the explained coincident condition; therefore, it is visible in the uncorrelated and the beam-off spectra.}
    \label{fig:2d-hist_b} 
  \end{subfigure}
  \caption{Different spectra showing the $^{136}$Ba data for detector $\#$5 (n-type COAX with 95$\%$ detection efficiency). In (a), one can see the different evolution of $\mu$X and $\gamma$ rays over time in the correlated spectrum. In (b), it is possible to see that evolution represented in different spectral projections.}
  \label{fig:2d-hist}
\end{figure*}
The experimental setup, measurement principle, and analysis procedures used for M\textsc{onument}'s 2021 measurement campaign were presented.

We introduced the use of two parallel DAQ systems -- MIDAS and ALPACA -- and described their advantages and disadvantages, motivating their use in complementary analyses.
The analysis procedures for both of these systems were presented, along with a comprehensive description of all pertinent performance parameters required for extracting the relevant observables in the experiment.

Highlighted is the diagnostic run for ALPACA, our in-house-made DAQ, where a significant reduction of the livetime was found as compared to MIDAS.
Mitigation measures were proposed, which will be used in future campaigns and are expected to make the livetimes of the two systems compatible.
The aspects where the offline signal processing characteristic from ALPACA is an advantage to MIDAS were also reported, such as in the case of the energy resolution.
Lastly, the experimental rates were reported, and it was noted that we still need further understanding of one of our counters' spectra, which could potentially improve the data quality and reduce the fraction of random coincidences.

Overall, it was shown that there is enough information and understanding of our data to perform the high-level analysis.
Fig.~\ref{fig:2d-hist_a} features a 2D histogram produced for the $^{136}$Ba data after applying the coincident trigger condition.
One can observe the intensity distribution of $\gamma$ and $\mu$X rays over time in the correlated spectrum.
The fit of the $\gamma$ rays' decay profile is what will allow the calculation of the total capture rates.

Fig.~\ref{fig:2d-hist_b} shows different spectral projections used to understand the data. 
The final measurements' results will be reported in the upcoming publications and are expected to advance future calculations of nuclear matrix elements relevant for \onbb decay searches.
The energy projection of the 2D histogram, which is the correlated spectrum, will be used for the extraction of the partial capture rates.

%% file: 7-acknowledgements.tex
\section{Acknowledgments}\label{sec:acknoledgments}
The reported study was funded by RFBR and DFG, project number 21-52-12040, the DFG Grant 448829699, by a Department of Energy Grant No. DE-SC0019261, by FWO-Vlaanderen (Belgium), and by BOF KU Leuven under contract No. C14/22/104.
The measurements were performed at the High-Intensity Proton Accelerator using the Swiss Infrastructure for Particle Physics at the Paul Scherrer Institute. 
The authors are grateful to the $\mathrm{\pi}$E1-2 beamline, technical support, muX and MIXE groups whose outstanding efforts have made these experiments possible.
I. Ostrovskiy thanks the Chinese Academy of Sciences (CAS) President's International Fellowship Initiative (PIFI) for the support.